\documentclass[twocolumn]{aastex62}
\usepackage[]{natbib,aas_macros}
\newcommand\Ha{H$\alpha$}

\newcommand\Oii{[\ion{O}{2}]}
\newcommand\rvir{r_{\mathrm{vir}}}
\newcommand\mssfr{\langle\mathrm{SSFR}\rangle}

\accepted{16 Jun, 2020}

\shorttitle{}
\shortauthors{Asano et al.}

\begin{document}

\title{Environmental impact on star-forming galaxies in a $z \sim 0.9 $ cluster during course of galaxy accretion}

\correspondingauthor{Tetsuro Asano}
\email{t.asano@astron.s.u-tokyo.ac.jp}

\author{Tetsuro Asano}
\affiliation{Department of Astronomy, University of Tokyo, 7-3-1 Hongo, Bunkyo-ku, Tokyo, 113-0033, Japan}
\affiliation{Astronomical Institute, Tohoku University, Aoba-ku, Sendai 980-8578, Japan}

\author{Tadayuki Kodama}
\affiliation{Astronomical Institute, Tohoku University, Aoba-ku, Sendai 980-8578, Japan}

\author{Kentaro Motohara}
\affiliation{Institute of Astronomy, University of Tokyo, 2-21-1 Osawa, Mitaka, Tokyo 181-0015, Japan}

\author{Lori Lubin}
\affiliation{Department of Physics, University of California, Davis, 1 Shields Avenue, Davis CA 95616, USA}

\author{Brian C. Lemaux}
\affiliation{Department of Physics, University of California, Davis, 1 Shields Avenue, Davis CA 95616, USA}

\author{Roy Gal}
\affiliation{Institute for Astronomy, University of Hawai'i, 2680 Woodlawn Drive, Honolulu, HI 96822, USA}

\author{Adam Tomczak}
\affiliation{Department of Physics, University of California, Davis, 1 Shields Avenue, Davis CA 95616, USA}

\author{Dale Kocevski}
\affiliation{Department of Physics and Astronomy, Colby College, Waterville, ME 04901, USA}

\author{Masao Hayashi}
\affiliation{National Astronomical Observatory of Japan, 2-21-1 Osawa, Mitaka, Tokyo 181-8588, Japan}

\author{Yusei Koyama}
\affiliation{Subaru Telescope, 650 N. Aohoku Pl, Hilo, HI 96720, USA}

\author{Ichi Tanaka}
\affiliation{Subaru Telescope, 650 N. Aohoku Pl, Hilo, HI 96720, USA}

\author{Tomoko L. Suzuki}
\affiliation{Astronomical Institute, Tohoku University, Aoba-ku, Sendai 980-8578, Japan}
\affiliation{National Astronomical Observatory of Japan, 2-21-1 Osawa, Mitaka, Tokyo 181-8588, Japan}

\author{Naoaki Yamamoto}
\affiliation{Astronomical Institute, Tohoku University, Aoba-ku, Sendai 980-8578, Japan}

\author{Daiki Kimura}
\affiliation{Astronomical Institute, Tohoku University, Aoba-ku, Sendai 980-8578, Japan}

\author{Masahiro Konishi}
\affiliation{Institute of Astronomy, University of Tokyo, 2-21-1 Osawa, Mitaka, Tokyo 181-0015, Japan}

\author{Hidenori Takahashi}
\affiliation{Institute of Astronomy, University of Tokyo, 2-21-1 Osawa, Mitaka, Tokyo 181-0015, Japan}
\affiliation{Kiso Observatory, the University of Tokyo, 10762-30, Mitake, Kiso-machi, Kiso-gun, Nagano 397-0101, Japan}

\author{Yasunori Terao}
\affiliation{Institute of Astronomy, University of Tokyo, 2-21-1 Osawa, Mitaka, Tokyo 181-0015, Japan}

\author{Kosuke Kushibiki}
\affiliation{Institute of Astronomy, University of Tokyo, 2-21-1 Osawa, Mitaka, Tokyo 181-0015, Japan}

\author{Yukihiro Kono}
\affiliation{Institute of Astronomy, University of Tokyo, 2-21-1 Osawa, Mitaka, Tokyo 181-0015, Japan}

\author{Yuzuru Yoshii}
\affiliation{Institute of Astronomy, University of Tokyo, 2-21-1 Osawa, Mitaka, Tokyo 181-0015, Japan}
\affiliation{Steward Observatory, University of Arizona, 933 North Cherry Avenue, Rm. N204 Tucson, AZ 85721-0065, USA}

\author{the SWIMS team}

\begin{abstract}
Galaxies change their properties as they assemble into clusters.
In order to understand the physics behind that, we need to go back in time and observe directly what is occurring in galaxies as they fall into a cluster.
We have conducted a narrow-band and $J$-band imaging survey on a cluster CL1604-D at $z=0.923$ using a new infrared instrument SWIMS installed at the Subaru Telescope.
The narrow-band filter, NB1261, matches to \Ha\ emission from the cluster at $z=0.923$.
Combined with a wide range of existing data from various surveys, we have investigated galaxy properties in and around this cluster in great detail.
We have identified 27 \Ha\ emitters associated with the cluster. They have significant overlap with MIPS 24$\micron$ sources and are located
exclusively in the star forming regime on the rest-frame $UVJ$ diagram. 
We have identified two groups of galaxies near the cluster in the 2D spatial distribution and the phase-space diagram, which are likely to be in-falling to the cluster main body.
We have compared various physical properties of star forming galaxies, such as specific star formation rates (burstiness) and morphologies (merger) as a function of environment; cluster center, older group, younger group, and the field.
As a result, a global picture has emerged on how the galaxy properties are altered as they assemble into a denser region. 
This includes the occurrence of mergers, enhancement of star formation activity, excursion to the dusty starburst phase, and eventual quenching to a passive phase.
\end{abstract}

\keywords{galaxies: clusters: individual (CL1604-D) -- galaxies: evolution -- galaxies: star formation}

\section{Introduction}\label{sec:intro}

\subsection{Environmental dependence of galaxies and distant cluster surveys}
It is well established, in particular in the local Universe, that galaxy properties, such as morphology and star forming activity,
strongly depend on surrounding environments \citep[e.g,][]{1980ApJ...236..351D}.
This trend is seen also at fixed stellar mass, meaning that some external effects specific to dense environment are required to shape galaxies \citep[e.g.,][]{2010ApJ...721..193P}.
At higher redshift, the relationships mentioned above seems to persist at least $z \sim 1.5$ \citep[e.g., ][]{2019MNRAS.490.1231L, 2019MNRAS.484.4695T, 2016A&A...592A.161N}. 
Although several physical processes have been proposed such as ram-pressure stripping and galaxy-galaxy interactions or mergers, the relative importance of these processes is yet unknown.
To reveal the physical mechanisms, it is crucial for us to go back in time and study distant clusters and their surrounding regions.
Mergers and interactions occur more frequently there, and we can directly see what physical processes are actually occurring in galaxies when they are falling into clusters.
This requires wide field observations to cover the in-falling regions around clusters.
There have been many systematic surveys so far conducted for distant clusters at $0.5<z<1.2$, such as Red-Sequence Cluster Survey-1 \citep[RCS-1;][]{2005ApJS..157....1G}, 
RCS-2 \citep{2011AJ....141...94G}, 
Panoramic Imaging and Spectroscopy ofCluster Evolution with Subaru \citep[PISCES;][]{2005PASJ...57..309K}, 
ESO Distant Cluster Survey \citep[EDisCS;][]{2005A&A...444..365W}, 
ACS Intermediate Redshift Cluster Survey \citep[ACSIRCS][]{2009ApJ...690...42M}, 
Observation of Redshift Evolution in Large Scale Environments \citep[ORELSE;][]{2009AJ....137.4867L}, 
Gemini CLuster Astrophysics Spectroscopic Survey \citep[GCLASS;][]{2012ApJ...746..188M}, 
Gemini Observations of Galaxies in Rich Early ENvironments \citep[GOGREEN;][]{2017MNRAS.470.4168B} 
and Massive and Distant Clusters of WISE Survey \citep[MaDCoWS;][]{2019ApJS..240...33G}.
We have much better understanding on when and where star forming galaxies are quenched as they arrive into clusters from their surrounding regions.
However, since the time scale of galaxy quenching appears rather short ($\lesssim$1Gyr), though perhaps occurring after a relatively long delay (see, e.g., \citealt{2019MNRAS.490.1231L} and references therein), the fraction of these galaxies is rather small,
at least to $z\lesssim1.5$, and is significantly less than that in the coeval field,
which makes it difficult to put definitive constraints on the main physical process or processes associated with their quenching.
Moreover, unlike older, passively evolving galaxies which can be easily recognized and thus securely selected as red sequence galaxies,
younger, star forming galaxies are much harder to select with high completeness {\it and} low contamination.
Previous studies tend to use optical SED or \Oii\ emission lines to characterize star forming activity. However these indicators are subject to severe dust attenuation which hide their intrinsic properties.
A better indicator of star formation that is often used and well calibrated in the local Universe is the hydrogen \Ha\ emission line, 
as it is relatively less affected by dust extinction due to its relatively longer wavelength (6563 \AA\ in the rest frame) than \Oii\ or rest-frame UV, although even some {\Ha} emissions are heavily attenuated as we show later.
At $z>0.5$, however, this useful line is redshifted to the near infrared, and thus limited specifications of instruments (cost, field of view, spectral multiplicity)
and brightness of the night sky (both thermal emission and OH sky lines) have been hampering its efficient observations.
Narrow-band imaging of \Ha\ emitters in a cluster, in between the strong OH sky lines, with a wide-field imager, is a good solution.
It also does not require pre-selection of targets unlike spectroscopic observations, and it can trace star forming galaxies
with \Ha\ flux density down to a certain observational limit.

\citet{2010MNRAS.403.1611K} actually did such a narrow-band \Ha\ emitter survey for a cluster (RXJ1716+67) at $z$=0.81 using Multi-Object Infrared Camera and Spectrograph \citep[MOIRCS;][]{2006SPIE.6269E..16I} on Subaru Telescope.
(see also \citealt{2004MNRAS.354.1103K} and \citealt{2011ApJ...734...66K} for $z=0.4$ clusters).
Together with AKARI \citep{2007PASJ...59S.369M, 2007PASJ...59S.401O} mid infrared imaging data, it revealed a void of star formation activities in the central core of the cluster,
and interestingly, it identified dusty star forming galaxies on or just below the red sequence in medium density regions and in-falling clumps.
This population seems to be related to the environmental effects. In fact, in-falling clumps are the sites where galaxy color distribution is drastically 
changed and where recently-quenched galaxies with strong Balmer absorption lines are seen in the stacked spectra at $z\sim0.8$. \citep{2006MNRAS.365.1392T}.
It is thus critical to investigate those galaxies in in-falling groups more in detail in order to understand the external environmental effects.

In this paper, we will focus on a rich cluster at $z$=0.9 \citep[CL1604-D;][]{2009AJ....137.4867L, 2011ApJ...736...38K, 2012ApJ...745..106L} and its surrounding region embedded in a gigantic supercluster
stretching over more than 100 Mpc in co-moving scale. This field has been extensively studied including Hubble Space Telescope (HST) imaging, Spitzer observations, and multi-object spectroscopy,
as summarized in the next section.
We have recently conducted a narrow-band \Ha\ emitter survey in and around this cluster with a brand-new wide-field NIR instrument Simultaneous-color Wide-field Infrared Multi-object Spectrograph \citet[SWIMS;][]{2016SPIE.9908E..3UM, 2018SPIE10702E..26K}
which is temporarily mounted on Subaru telescope. In this paper, we report its first results and discuss how the star formation activities are affected as the
galaxy groups are falling into the cluster by taking the full advantage of the wealth of data available for this cluster embedded in the supercluster.
The structure of this paper is the following: \S2 will summarize the properties of the supercluster from the previous work already presented
in the literature. \S3 will show the rich observational data sets consisting of both the existing data and the newly obtained data, and will also describe the data reduction.
Main analyses of the data to derive physical quantities will be shown in \S4, and the results will be presented in \S5. Discussion follows in \S6 and the
summary will be given in \S7.
All the magnitudes presented in this paper are AB magnitudes.
We adopt cosmology with $H_0=70\; \mathrm{km\; s^{-1} \; Mpc^{-1}}$, $\Omega_{\mathrm{m}} = 0.3$ and $\Omega_{\Lambda}=0.7$.

\section{The SC1604 supercluster}

\begin{figure}
	\centering
	\includegraphics[width=\linewidth]{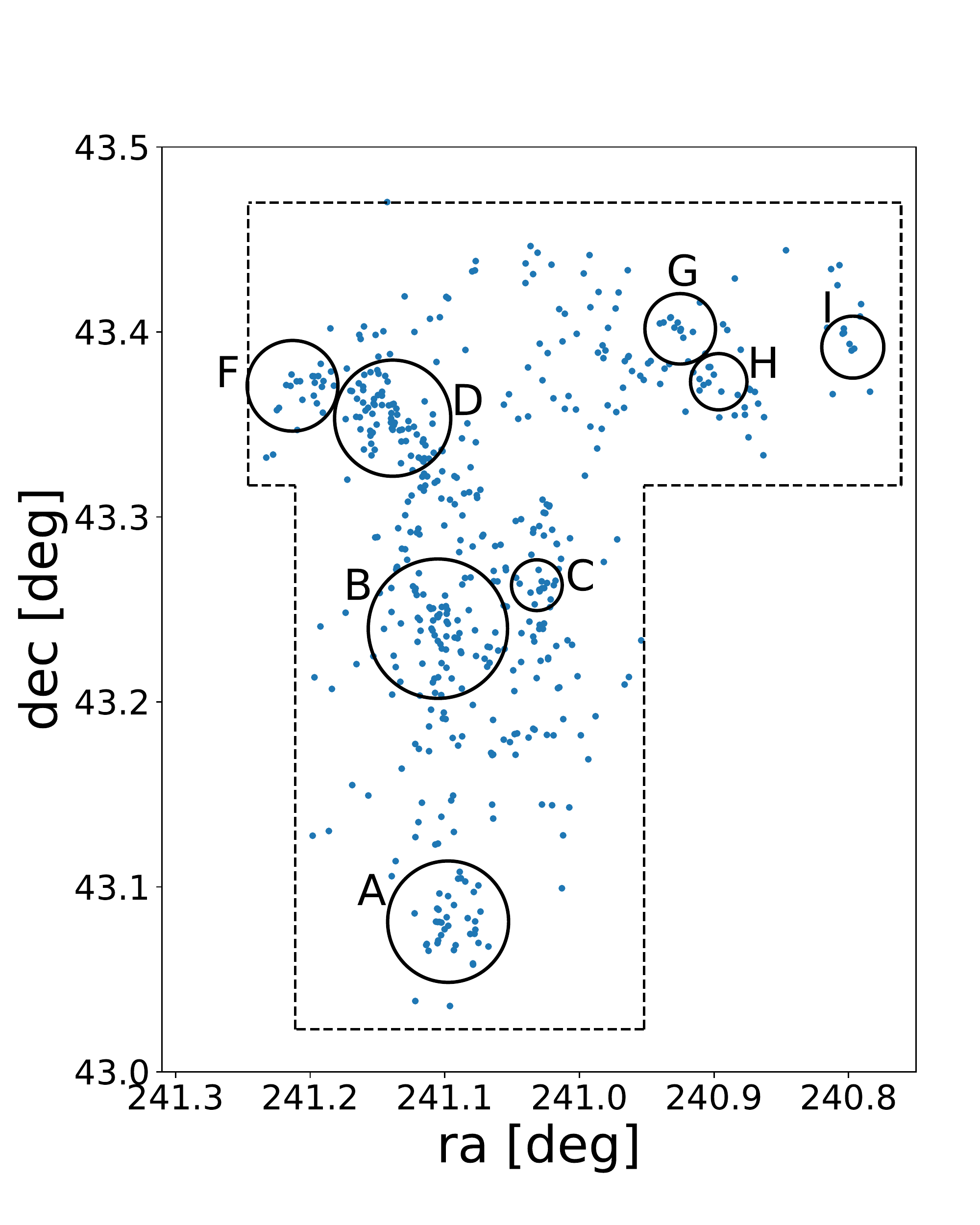}
	\caption{Spatial distribution of the spectroscopically-confirmed galaxies in SC1604 supercluster. Galaxies whose redshifts are between 0.84 and 0.96 are plotted. The circles denote the virial radius of each cluster/group. The dashed box indicates the region of the spectroscopic observation.\label{fig:supercluster_spatial}}
\end{figure}

The SC1604 supercluster is a known large scale structure at $z \sim 0.9$, 
and has been observed as a part of the ORELSE survey \citep{2009AJ....137.4867L, 2008ApJ...684..933G}.
It is one of the most widely studied high-redshift superclusters observed from various perspectives such as optical and NIR imaging, optical spectroscopy, MIR imaging, HST imaging, and X-ray observations. 

We show in Figure~\ref{fig:supercluster_spatial} a spatial distribution of galaxies in the supercluster and virial radii of identified clusters (A, B and D) and groups (C, F, G, H and I), reproduced from \citet{2012ApJ...745..106L} and \citet{2017MNRAS.472.3512T}. 
They identified $\sim500$ member galaxies of SC1604 supercluster from the spectroscopy with Keck~I/Low-Resolution Imaging Spectrometer \citep[LRIS;][]{1995PASP..107..375O} and Keck~II/DEep Imaging Multi-Object Spectrograph \citep[DEIMOS;][]{2003SPIE.4841.1657F}.
The spectroscopy was designed to target the red-sequence overdensities identified in \citet{2008ApJ...684..933G} as well as the surrounding filamentary structure, with particularly dense coverage in and around the target of this study, the CL1604-D cluster. See \citet{2012ApJ...745..106L} for more details on the spectroscopy.
In the figure, 517 supercluster members, whose spectral redshifts are between 0.84 and 0.96, are plotted. It spans $\sim 10$~Mpc on the sky and is one of the most widely mapped structures at high redshifts ($>0.5$)
Other examples are  RCS2319+00 at $z\sim0.9$ \citep{2008ApJ...677L..89G}; supercluster at $z\sim0.91$ in the SSA22 field \citep{2016ApJ...821L..10K} and Lynx supercluster at $z\sim1.3$ \citep{1999AJ....118...76R,2005MNRAS.357.1357N}.

Furthermore, the recent wide-field deep imaging survey by the Subaru/Hyper Suprime-Cam~\citep[HSC;][]{2018PASJ...70S...1M, 2018PASJ...70S...2K, 2018PASJ...70...66K, 2018PASJ...70S...3F} suggests the large scale structure expands even wider than the area used previously to map SC1604 \citep{2019PASJ...71..112H}. 
It seems to spans $\sim50$~Mpc scale.

Two X-ray detected clusters (cluster A and B) are dominant in SC1604, and smaller systems surround them. Cluster D is the third most massive system, which is also X-ray detected at lower $L_{X}$ (see \citealt{2018MNRAS.478.1403R}), appears dynamically younger than the other two massive clusters. The fraction of star-forming galaxies in this structure is higher than either of the other massive clusters, including many active dusty starbursts, and there are few signs of recent widespread quenching activity among its members \citep{2011ApJ...736...38K, 2012ApJ...745..106L, 2014ApJ...792...16W}.
Details of previous observations and studies about SC1604 supercluster are described in \citet{2004ApJ...607L...1G}, \citet{2008ApJ...684..933G}, \citet{2009ApJ...690..295K, 2009ApJ...700..901K, 2011ApJ...736...38K}, \citet{2010ApJ...716..970L, 2012ApJ...745..106L}, \citet{2014ApJ...792...16W} and \citet{2017MNRAS.472.3512T, 2019MNRAS.484.4695T}.

\subsection{Our Target: CL1604-D}
\begin{figure}
	\centering
	\includegraphics[width=\linewidth]{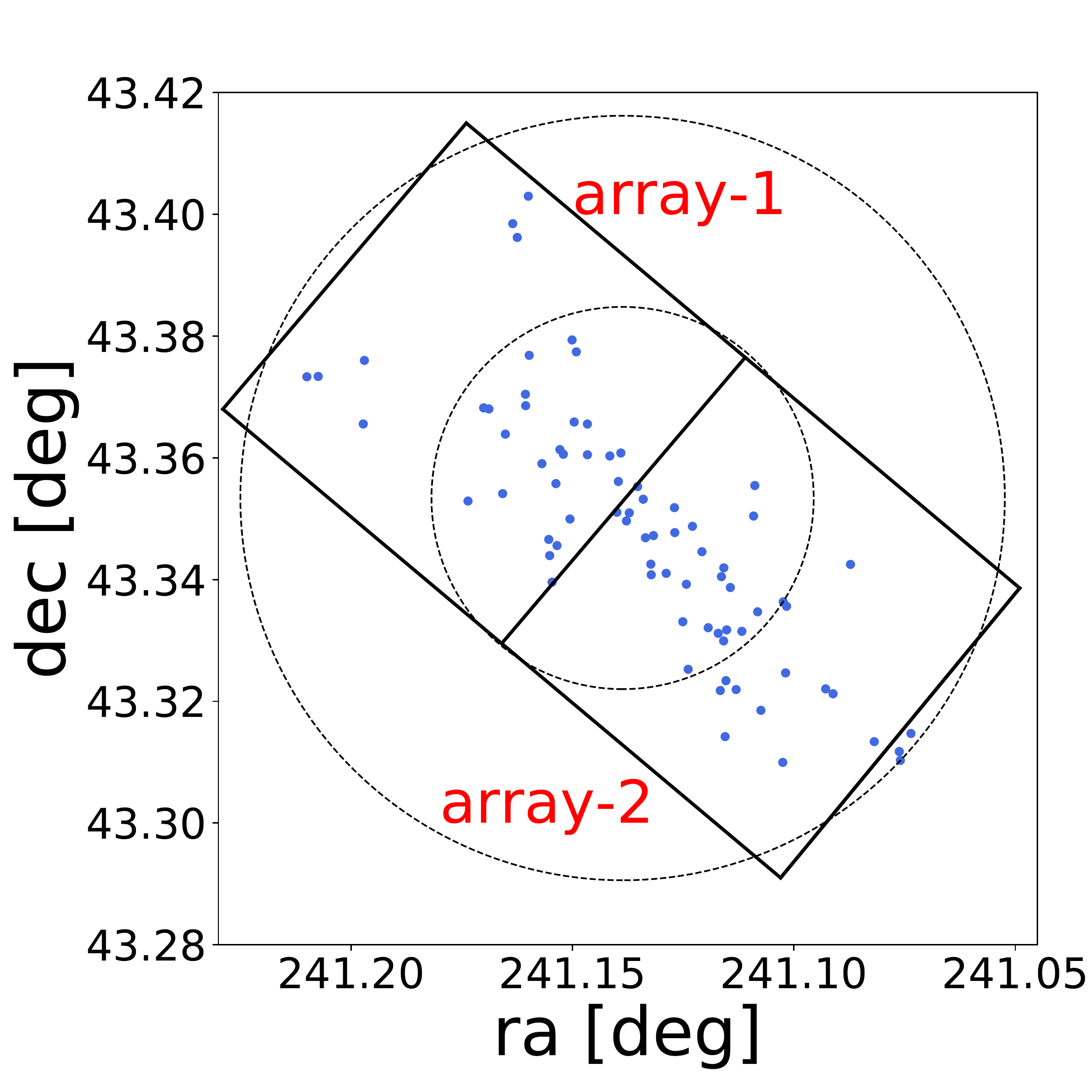}
	\caption{Spatial distribution of CL1604-D member galaxies and the FoV of SWIMS.\@ 
	Two circles denote the virial radius of the cluster and the twice of that value, respectively. Two squares denote the pointings FoV's of the two arrays of SWIMS used in our observation.\label{fig:spatial_dist_and_fov}}
\end{figure}

Our target, CL1604-D cluster, is the third most massive cluster in the SC1604 supercluster system, whose velocity dispersion and median redshift are $\sigma = 688 \pm 88 \; \mathrm{km \; s^{-1}}$ and $z=0.923$ respectively \citep{2018MNRAS.478.1403R, 2020MNRAS.491.5524H}. The velocity dispersion can be converted to $r_{200}$: the radius within which the average density is 200 times the critical density of the universe at the redshift of the system, by the equation of \citet{1997ApJ...478..462C}:
\begin{equation}
	r_{200} =\frac{\sqrt{3}\sigma}{10H(z)}
	\label{eq:r200}
\end{equation}
where $H(z)$ is the Hubble parameter at the redshift of the cluster. 
The relation between $r_{200}$ and $\rvir$, the virial radius of the system, is given by $\rvir = r_{200}/1.14$ \citep{2006A&A...456...23B,2009ApJ...693..112P}.
The determined value of $\rvir$ for the cluster is 0.89~Mpc.
In our study, we select a galaxy physically associated to the cluster and its surrounding structures as those which satisfies 
(1) $r<2\rvir$ where $r$ is projected distance from the center, 
 and (2) $|\Delta v| < 3\sigma$ where $\Delta v$ is a line-of-sight velocity measured from the median redshift of the cluster. 
We adopt the cluster center defined in \citet{2008ApJ...684..933G}. They determine it as the centroid of the density of red galaxies.
 We note that it is very close ($\lesssim 100$ $h_{70}^{-1}$ kpc) to other central coordinates in \cite{2018MNRAS.478.1403R}, who determine X-ray, and spectral member luminosity-/stellar-mass-weighted centers.
The spatial distribution of the cluster members is shown in Figure~\ref{fig:spatial_dist_and_fov}. The galaxy distribution clearly shows a filamentary structure, and, as discussed in the previous studies and the later section, this system is not yet dynamically relaxed and still in the process of galaxy accretion.

\section{data}\label{sec:data}
\subsection{NIR Imaging}\label{sec:nir_imaging}
\subsubsection{Observations}\label{sec:obs}

\begin{figure}
	\centering
	\includegraphics[width=\linewidth]{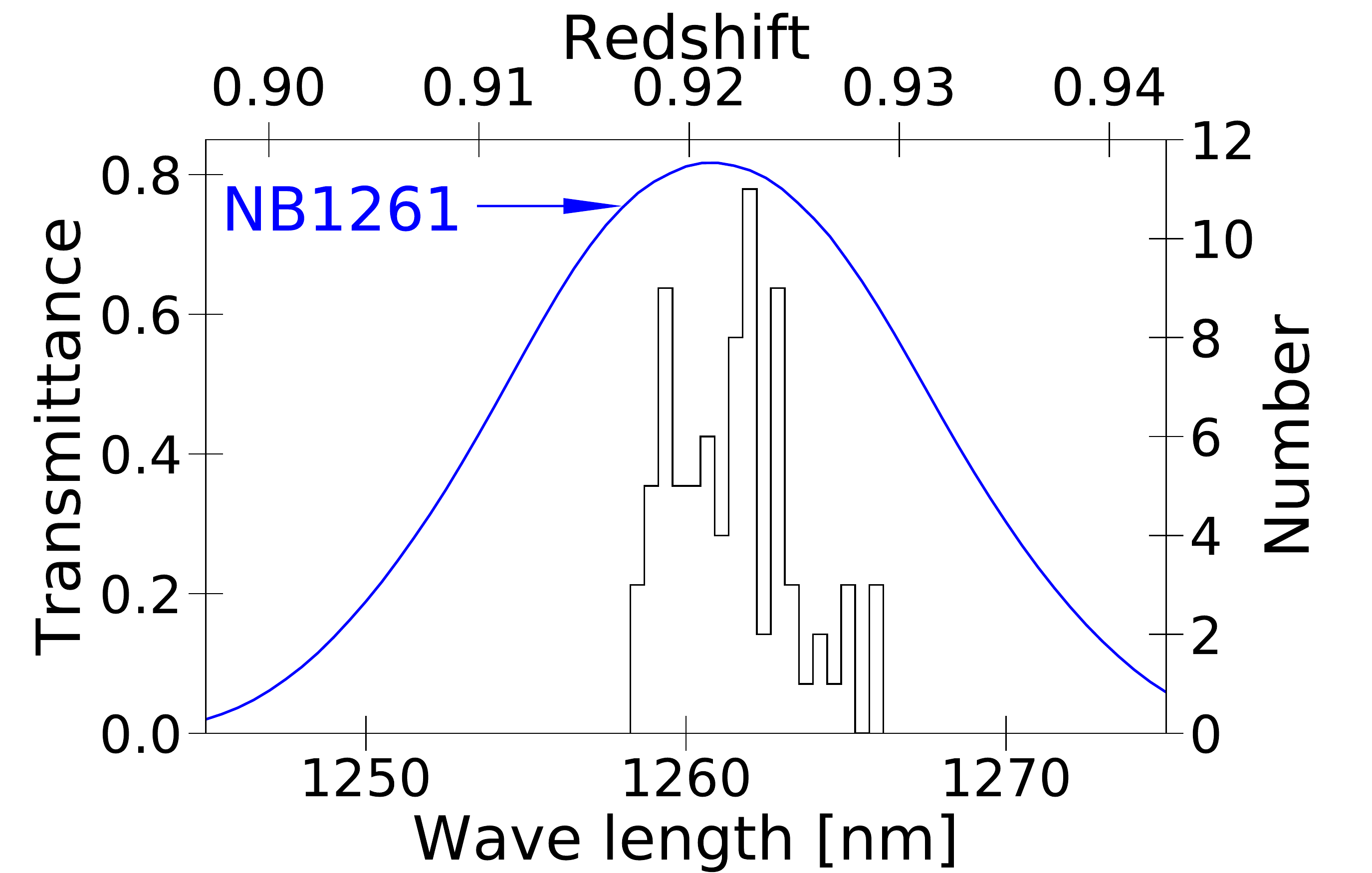}
	\caption{Response function of NB1261 filter and redshift distribution of CL1604-D members. Top axis shows the redshift range and the bottom axis shows the corresponding observed wavelength of the {\Ha} line for each redshift.\label{fig:response_and_z_dist}}
\end{figure}

In the observations of CL1604-D cluster, we used SWIMS which is a new instrument developed for the University of Tokyo Atacama Observatory (TAO) 6.5m telescope \citep{2016SPIE.9906E..0RY}, being constructed on the summit of Co.\ Chajnantor in Chile. It is designed for simultaneous imaging or spectroscopy in two wavelength channels, 0.9--1.4 and 1.4--2.5 {\micron}.
The overview of the instrument is described in \citet{2016SPIE.9908E..3UM} and \citet{2018SPIE10702E..26K}.
Before commissioning at the TAO 6.5m telescope in Chile, it has been installed on Subaru Telescope for engineering observations. As a part of such engineering runs to test the deep imaging performances, we have observed CL1604-D on 2018 May 31 and 2019 January 24.
The field of view (FoV) of the observations is shown in Figure~\ref{fig:spatial_dist_and_fov}. FoV and pixel scale are $6.6' \times 3.3'$ and 0.096$^{"}$/pixel, respectively, on Subaru Telescope. Each focal plane is covered by two HAWAII-2RG arrays, and two squares in Figure~\ref{fig:spatial_dist_and_fov} denote them.
The imaging observations were performed in two filters: $J$ (central wavelength  $\lambda_{\mathrm{c}}=1.25$ \micron; FWHM $\Delta \lambda=0.16$ \micron) and NB1261 ($\lambda_{\mathrm{c}}=1.261$ \micron; $\Delta \lambda=0.012$ \micron).
The transmittance curve of the NB1261 filter and a redshift distribution of cluster members in CL1604-D are shown in Figure~\ref{fig:response_and_z_dist}. The peak of the filter transmittance neatly matches with the redshift distribution of the cluster members for their {\Ha} line. Therefore, we can trace {\Ha} emission line from all the galaxies in the cluster with this narrow-band filter.

\begin{deluxetable*}{cccccc}
	\tablecaption{Summary of the observation and photometry.\label{tab:obs_summary}}
	\tablehead{
		\colhead{Band ($\lambda_{\mathrm{c}}$/\micron, $\Delta \lambda$/\micron)} & \colhead{Obs. date} & \colhead{Exp. time (min.)} 
									 & \colhead{Seeing (arcsec.)} 
									 & \colhead{$3\sigma$ limiting mag. (array-1/array-2)} 
	}
	\startdata
		$J$ (1.25, 0.16) & May 31, 2018 & 27 &  0.60 & 23.78/23.72\\
		NB1261 (1.261, 0.012)& Jan. 24, 2019 & 54 & 0.65 & 23.48/23.43 
	\enddata
\end{deluxetable*}

\subsubsection{{\Ha} Image Reduction}
The SWIMS data reduction was carried out using the SWSRED software. It is based on MCSRED2\footnote{\url{http://www.naoj.org/staff/ichi/MCSRED/mcsred.html}}, a pipeline software written by Ichi Tanaka for MOIRCS on Subaru Telescope. After image cleaning, position matching and frame co-adding by SWSRED, we estimated seeing size by measuring the FWHM of point spread functions (PSF) in the combined image with IRAF\footnote{\url{http://iraf.noao.edu}} as indicated in Table 1.
Object detections and photometry in the $J$- and NB1261 bands were carried out with ``double-image mode'' of SExtractor \citep{1996A&AS..117..393B}. We use ``MAG\_APER'' with an aperture diameter of twice seeing size ($1\farcs3$) for $J-$NB1261 color and ``MAG\_BEST'' for individual $J$ and NB1261 magnitudes.
Magnitude zero-point in the $J$ band was determined by the Two-Micron All Sky Survey \citep[2MASS;][]{2006AJ....131.1163S} stars in the observed FoV.
We derived the NB1261 zero-point so that the $J-$NB1261 colors of bright galaxies (which are unlikely to be emitters) become zero.
Specifically, the median $J-$NB1261 color for galaxies with $J=15 - 19$ were adjusted to zero. 
1$\sigma$ errors of this fitting are 0.04 mag and 0.03 mag for array-1 and array-2, respectively. They lead $\sim 5$\% uncertainty in a line flux.
Finally, we estimated limiting magnitudes in each band and in each detector array from the sky flux dispersion measured by random aperture tests: We performed $1\farcs3$ aperture photometry at randomly selected 1000 positions in the FoV.
$3\sigma$ limiting magnitudes are $\sim23.7$ mag in the $J$-band and $~23.4$ mag in the NB1261 band, respectively. 
These magnitudes correspond to $\sim 4 \times 10^9M_{\sun}$ for passively evolving galaxies on the red sequence. It can reach down to $\sim 2 \times 10^9 M_{\sun}$ for star-forming galaxies.
We show a summary of our observations and photometry in Table~\ref{tab:obs_summary}.

\subsection{Optical Imaging}
The SC1604 supercluster is located in the survey area of the ``Wide'' layer of Hyper Suprime-Cam Subaru Strategic Program \citep[HSC-SSP;][]{2018PASJ...70S...4A, 2018PASJ...70S...8A}. It is an ongoing legacy imaging survey program with HSC \citep{2018PASJ...70S...1M, 2018PASJ...70S...2K, 2018PASJ...70...66K, 2018PASJ...70S...3F} on Subaru Telescope, covering the unprecedentedly large areas of the sky for 8-10m class telescopes. The large FoV of HSC~($1.77 \; \deg^2$) enables us to build a gigantic sample of galaxies in various fields which can be used for many science topics in modern astronomy and cosmology.
The vast amount of data obtained in this survey were reduced with hscPipe 
\citep{2018PASJ...70S...5B}, a pipeline software based on Large Synoptic Survey Telescope (LSST) stack framework \citep{2008SerAJ.176....1I, 2010SPIE.7740E..15A}, and photometric and astrometric calibrations for the data were performed with the Panoramic Survey Telescope And Rapid Response System (Pan-STARRS1) data \citep{2013ApJS..205...20M, 2012ApJ...756..158S, 2012ApJ...750...99T}.

In this study, we use data from the internal data release S17A \citep{2019PASJ...71..114A}, and our target supercluster exists in the HSC-Wide layer where imaging observations in $g$, $r$, $i$ , $z$ and $y$ filters are carried out to the depth of $\sim26$ mag.
In addition to the photometric catalog (magnitudes and colors), photometric redshifts and other physical quantities of individual galaxies computed by some photo-$z$ codes are also available. We adopt the stellar masses and the amount of dust attenuation estimated by MIZUKI code \citep{2015ApJ...801...20T, 2018PASJ...70S...9T}. It is based on an SED template fitting technique with Bayesian priors incorporated in order to provide us with physical properties of galaxies in realistic ranges.
It assumes Chabrier IMF \citep{2003PASP..115..763C},  Calzetti extinction curve \citep{2000ApJ...533..682C}, exponentially decaying star formation rates and solar metallicity. 
As parameters of stellar population synthesis, it uses ages between 0.05 and 14 Gyr with a step of $\sim 0.05$ dex, timescales of SFR ($\tau$) between 0.1 and 11 Gyr with a step of 0.2 dex in addition to $\tau=0$ and $\infty$, and optical depths in $V$ band ($\tau_V$) between 0 and 2 with a step of 0.1 in addition to $\tau_V=$ 2.5, 3, 4 and 5.
We also adopt Chabrier IMF \citep{2003PASP..115..763C} and Calzetti extinction curve \citep{2000ApJ...533..682C} in our analysis for consistency.
\subsection{Existing Data}
In addition to the {\Ha} imaging with SWIMS and data from HSC-SSP survey, we use data obtained in the ORELSE survey. ORLESE observations and data are summarized in \citet{2017MNRAS.472.3512T,2019MNRAS.484.4695T} and \citet{2019MNRAS.482.3514P}. 
Between the catalogs, nearest sources in $1\farcs0$ radius are identified as a same object.

\subsubsection{Optical Spectroscopy}
Spectroscopic redshift data used in this study were obtained by observations with DEIMOS \citep{2003SPIE.4841.1657F} on Keck II telescope. Targets for the spectroscopy were selected based on observed frame colors. Photometric samples whose observed frame $r'-i'$ and $i'-z'$ colors were consistent with the supercluster redshift ($z \sim 0.9$) were preselected. Furthermore, objects fainter than $i'=24.5$ mag were generally avoided as target.
As mentioned earlier, spectroscopic observations in and around cluster D were particularly dense, resulting in a 84\% of objects with F814W$<24$ and $r<2\rvir$ having a secure spectroscopic redshift. See \citet{2012ApJ...745..106L} for details of the target selection.
Data were reduced using a customized version of the Deep Evolutionary Extraglactic Probe2 \citep[DEEP2;][]{2003SPIE.4834..161D, 2013ApJS..208....5N} spec2d software. Redshifts were measured with DEEP2 zspec pipeline and assigned quality codes ($Q$). We use only redshifts with $Q \geq 3$, which indicate secure extragalactic objects. In addition, spectra of some galaxies in the cluster were obtained with LRIS \citep{1995PASP..107..375O} on Keck I telescope.
The number of total spectroscopically confirmed member which satisfy $r<2\rvir$ and $|\Delta v| < 3\sigma$ is 87.

\subsubsection{MIR Imaging}
The SC1604 supercluster was observed with the Spitzer Space Telescope as a part of the Spitzer program GO-30455 (PL: L. M. Lubin). We measure dust emission from the 24 {\micron} imaging with Multiband Imaging Spectrometer \citep[MIPS;][]{2004ApJS..154...25R}. We use data reduced by \citet{2019MNRAS.484.4695T}. They downloaded the data from the Spitzer Heritage Archive\footnote{\url{http://sha.ipac.caltech.edu/applications/Spitzer/SHA/}}, and performed the reduction using the MOPEX software package \citep{2006SPIE.6274E..0CM}: background matching between overlapping frames and mosaicing were performed using {\it overlap.pl} and {\it mosaic.pl} respectively, and the uncertainty per pixel were measured with {\it t-photo}.
The depth of MIPS data were estimated to be $\sim$40 $\mu$Jy by $10\farcs4$ photometry at 1000 randomly selected points. 

\subsubsection{HST imaging}
SC1604 supercluster was observed with the HST/Advanced Camera for Survey \citep[ACS;][]{1998SPIE.3356..234F} in F606W and F818W bands (GO-11003; PI:\ L. M. Lubin). 
Average integration time for CL1604-D is 4840s, which corresponds to $5\sigma$ limiting magnitudes of 28.1 and 27.6 mag in F606W and F818W, respectively. Details on the HST observations and reductions are presented in \citet{2009ApJ...700..901K}.

We use the HST/ACS imaging to select galaxies that are undergoing mergers/interactions and follow the morphological classification described in \citet{2011ApJ...736...38K}. We classify ongoing mergers or those likely to merge into four categories: merge (M:\ typically disturbed with an irregular companion),  interaction (I:\ these often exhibit tidal features and have a nearby companion), tidal disruption (T:\ these exhibit tidal features and do not have an obvious companion), 
or asymmetric morphology (A:\ possibly disturbed, these have asymmetric light distributions).
In this study, we define merger/interaction galaxies as those classified M, I or T in as done \citet{2011ApJ...736...38K}.

\subsubsection{SED Fitting}
\citet{2017MNRAS.472.3512T, 2019MNRAS.484.4695T} performed SED fitting on these photometry in order to derive photometric redshifts and rest-frame fluxes.
Existing data include photometry in the following bands: $B$, $V$, $R_C$, $I_C$ and $ Z_+$ filters with Suprime-Cam \citep{2002PASJ...54..833M} on Subaru Telescope; $r'$, $i'$ and $z'$ filters with the Large Format Camera \citep[LFC;][]{2000AAS...196.5209S} on the Palomar 200-inch Hale telescope; $J$ and $K$ filers with the Wide Field Camera \citep[WFCAM;][]{2007A&A...467..777C} on the United Kingdom InfraRed Telescope (UKIRT); [3.6], [4.5], [5.8] and [8.0] filters with the InfraRed Array Camara \citep[IRAC;][]{2004ApJS..154...10F} on Spitzer Space Telescope. 
They utilized Easy and Accurate Redshifts from Yale~\citep[EAZY;][]{2008ApJ...686.1503B}. EAZY estimates redshift and rest-frame flux simultaneously, but they provided redshift information in advance for objects whose spectroscopic redshifts are available. In this study, we use rest-frame colors for $U-V$ versus $V-J$ diagram ($UVJ$ diagram) and 2800 {\AA} luminosity~($l_{\nu,2800}$) from their fitting results. 
Combined with MIPS data, $l_{\nu,2800}$ is used to calculate SFR(FIR+UV).
Photometric redshifts estimated by MIZUKI and EAZY agree with each other within 0.1 for most of the galaxies in the cluster, and we confirm the consistency between the two SED fitting codes.

\section{Analysis}
\subsection{Selection of {\Ha} Emitters}\label{sec:emitter_select}

\begin{figure}
	\centering
	\includegraphics[width=\linewidth]{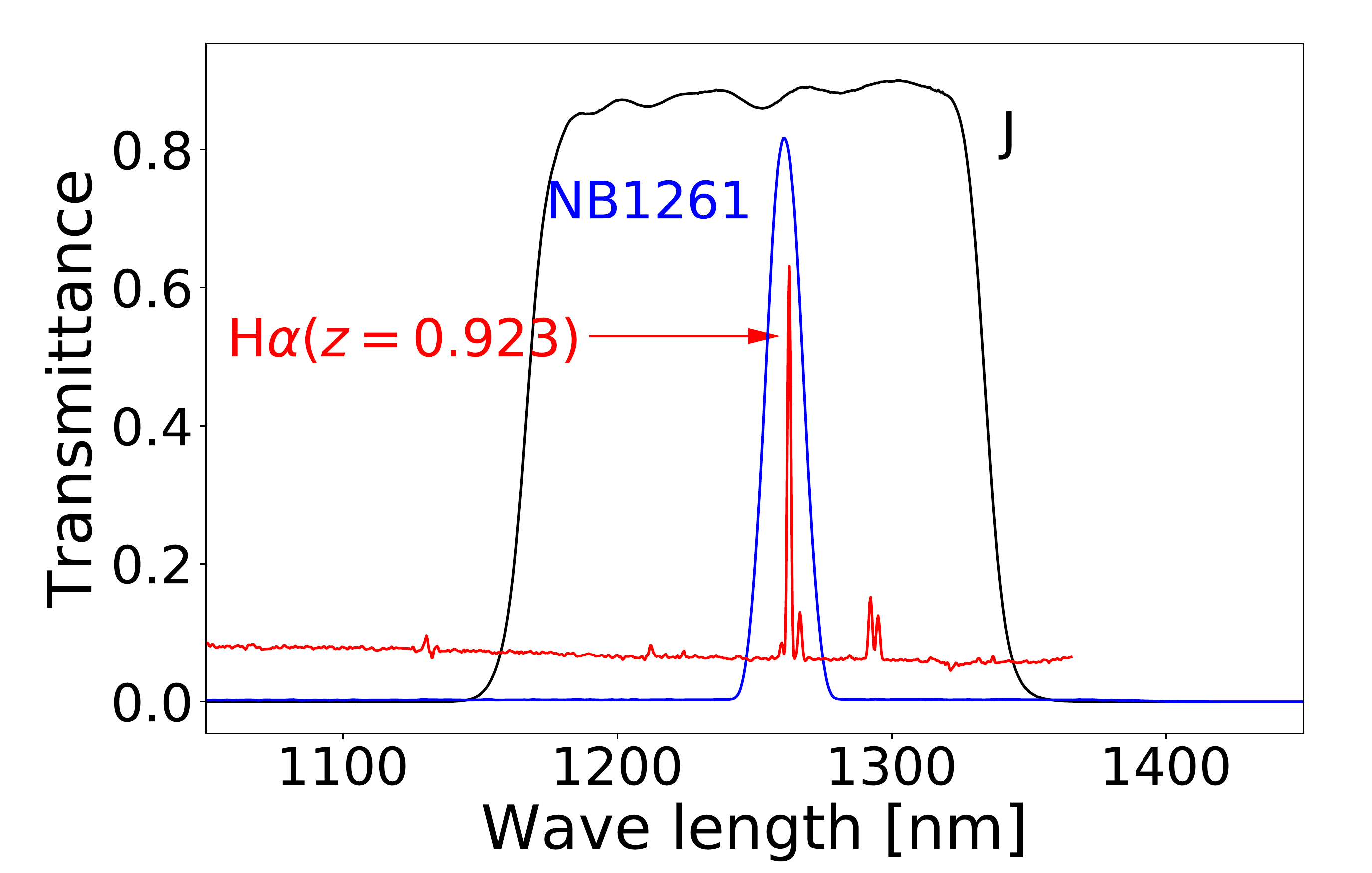}
	\caption{A template spectrum of HAE and the emittance curves of $J$ and NB1261 filters. We shift the spectrum of NGC4449 \citep{1992ApJS...79..255K} to $z=0.923$.\label{fig:respoce_func_and_template_SED}}
\end{figure}

\begin{figure*}
	\gridline{
		\fig{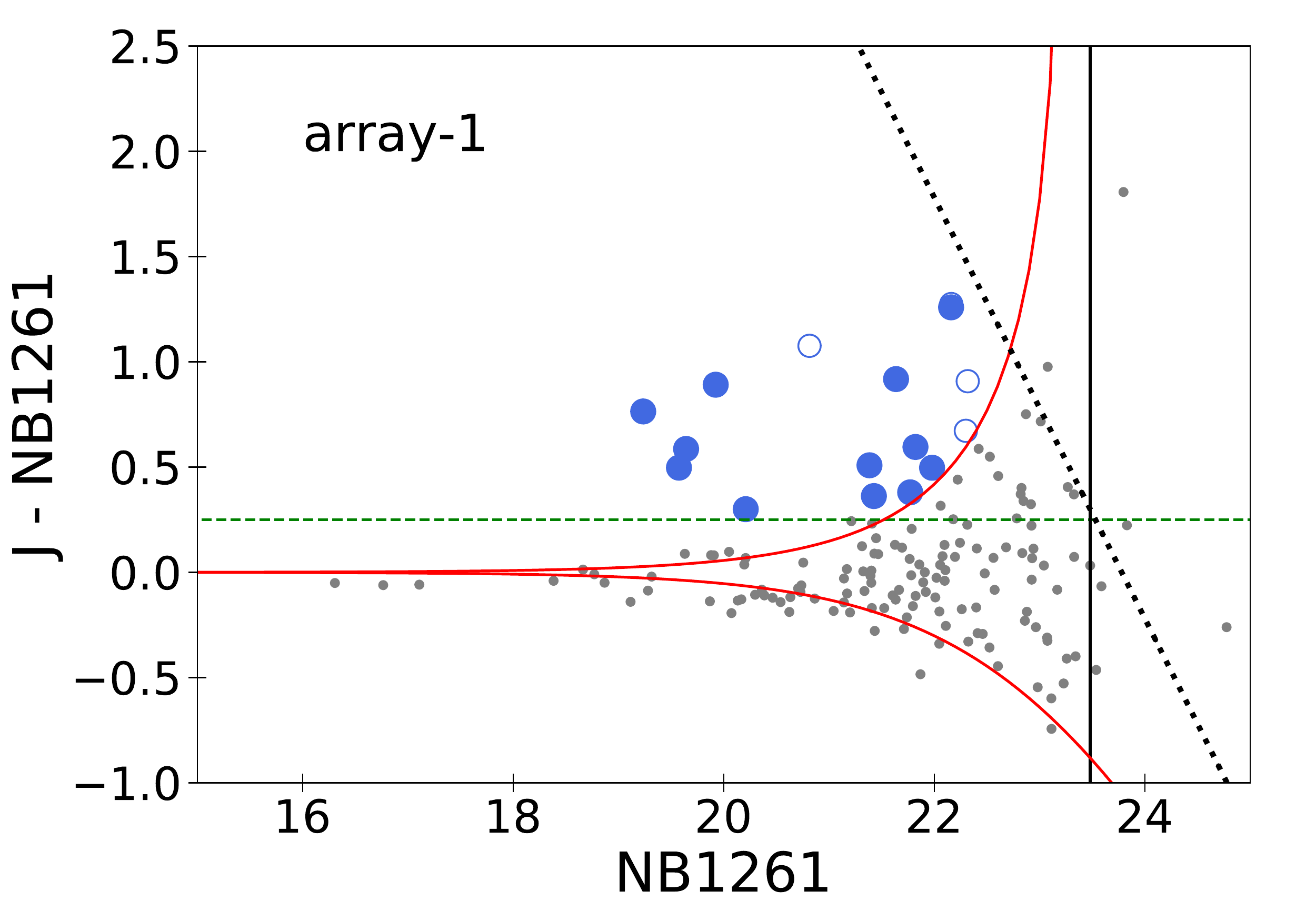}{0.5\linewidth}{}
		\fig{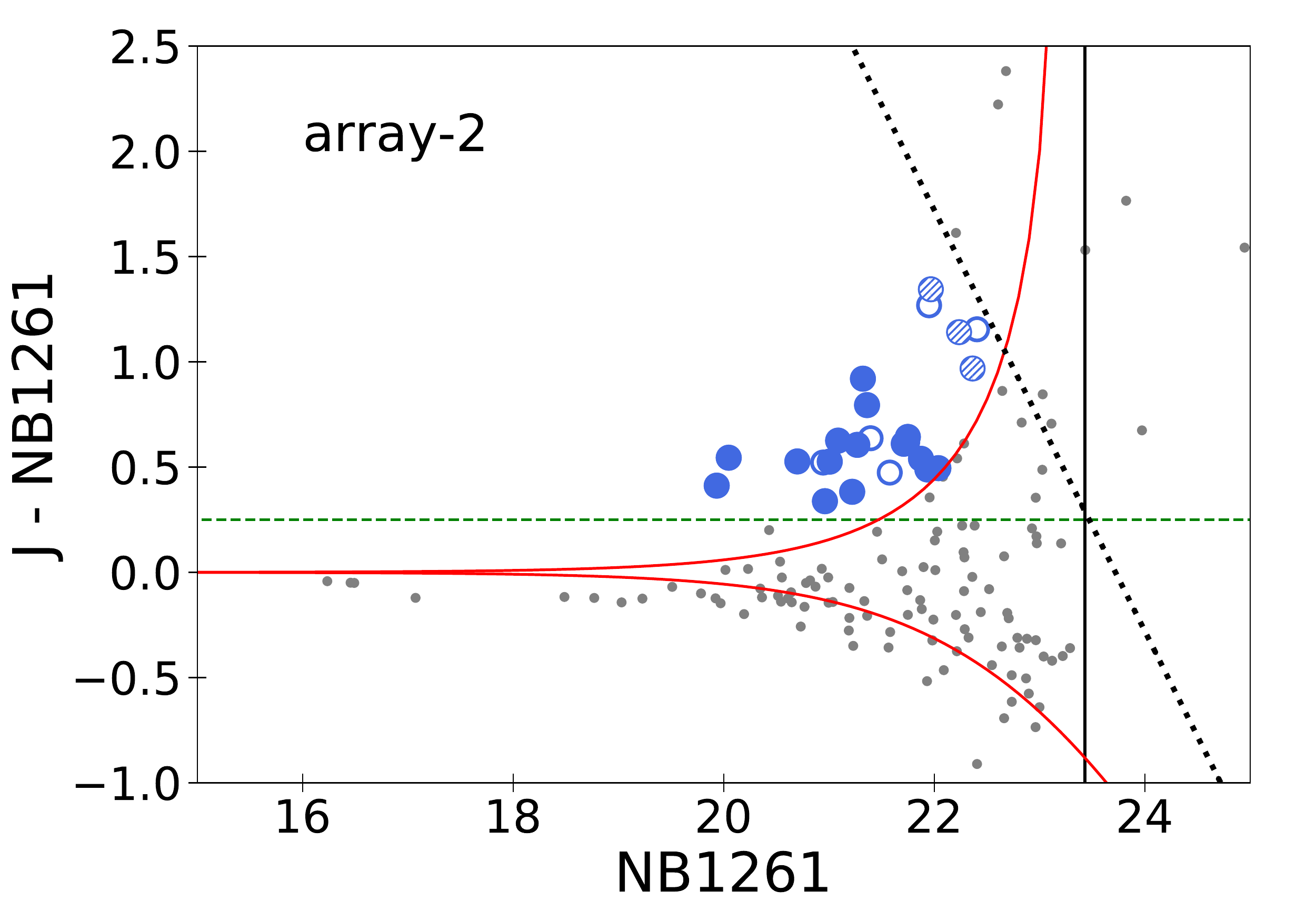}{0.5\linewidth}{}
	}
	\caption{$J-\mathrm{NB1261}$ versus NB1261 color-magnitude diagrams to select NB1261 emitters for each array (left: array-1, right: array-2). In both panels, solid and dotted lines show $3\sigma$ limiting magnitudes in NB1261 and $J$ respectively. Red curves denote $\pm 3\sigma$ errors in the $J-\mathrm{NB1261}$ color. Dashed line represents a color cut of $J-\mathrm{NB1261}=0.25$, which corresponds to equivalent width of 1.78~nm in the rest frame at $z=0.923$. NB1261 emitters are marked with blue circles, and filled ones represent spectroscopically confirmed HAEs. Emitters which have photo-$z$ of $zsim1.5$ are shown as hatched circles.\label{fig:emitter_selection} }
\end{figure*}
We show the response functions of $J$ and NB1261 filters in Figure~\ref{fig:respoce_func_and_template_SED}. NB1261 band lies in $J$ band, hence we can select emitters comparing magnitudes in the two bands.
If a emission line falls in the narrow band like Figure~\ref{fig:respoce_func_and_template_SED}, the flux density in the narrow band will be larger than that in the broad band. In other words, $J- \mathrm{NB1261}$ colors of emitters will be redder than those of non-emitters.
We show in Figure~\ref{fig:emitter_selection} the color-magnitude diagrams used to select NB1261 emitters. We define the NB1261 emitters as galaxies that satisfy the following conditions.
(1)	Both $J$ and NB1261 magnitudes are brighter than $3\sigma$ limiting magnitude.
(2)	$J-\mathrm{NB1261} > 3\Sigma$, where $\Sigma$ is $1\sigma$ error in $J-\mathrm{NB1261}$ color \citep{1995MNRAS.273..513B}.
(3)	$J-\mathrm{NB1261} > 0.25$. This color cut corresponds to equivalent width of 1.78~nm in the rest frame at $z=0.923$.
This third criterion removes non-emitters whose intrinsic $J-\mathrm{NB1261}$ colors scatter around zero due to their spectral shapes. We choose a specific value of 0.25, since some objects below the lower red curve ($-3\sigma$ error in $J- NB1261$ color) can have
$J-\mathrm{NB1261}$ colors down to $\sim$ $-$0.25 as seen in Figure~\ref{fig:emitter_selection}.
NB1261 emitters are marked with blue circles in the figure.
An NB1261 emitter is not necessarily an {\Ha} emitter (HAE) at $z=0.923$. It may possibly be another line emitter at another redshift such as an [\ion{O}{3}] emitter at $z=1.5$.
Therefore, in order to remove contamination, we extract only the galaxies from NB1261 emitters whose spectroscopic redshifts lie in the redshift range of CL1604-D cluster. 
Finally, we have 27 HAEs (blue filled circles in Figure~\ref{fig:emitter_selection}) in the cluster.
Three NB1261 emitters' spec-$z$ are $z \sim 0.9$ but slightly outside the redshift range of the cluster.
The other NB1261 emitters which are not selected as HAEs do not have counterparts in the spec-$z$ catalog.
Most of their photometric redshifts computed by the MIZUKI code are $z \sim 0.9$  except three having $z \sim 1.5$.

\subsection{Star Formation Rate}
In order to calculate star formation rates (SFRs), we first derive {\Ha} line fluxes.
Using $f_J$ and $f_{\mathrm{NB}}$, which are flux densities in $J$ and in NB1261 respectively, a line flux density, a continuum flux density and a rest-frame equivalent width are derived by the equations:
\begin{equation}
	F_{\mathrm{H}\alpha + \mathrm{[N \, II]}} =
	\Delta _{\mathrm{NB}}\frac{f_{\mathrm{NB}} - f_{J}}{1 - \Delta_{\mathrm{NB}}/\Delta_J}
	\label{eq:F_Ha}
\end{equation}																																						
\begin{equation}
f_{\mathrm{c}} = \frac{f_J-f_{\mathrm{NB}}(\Delta_{\mathrm{NB}}/\Delta_J)}{1-\Delta_{NB}/\Delta_J}
\label{eq:fc}
\end{equation}
\begin{eqnarray}
	 &\mathrm{EW}&_{\mathrm{rest}}(\mathrm{H}\alpha + [\mathrm{N\,II}])
	= {(1+z)}^{-1}\frac{F_{\mathrm{H}\alpha + [\mathrm{N\,II}]}}{f_{\mathrm{c}}} 
	\label{eq:EW}
\end{eqnarray}
where $\Delta_J$ and $\Delta_{\mathrm{NB}}$ are FWHMs of $J$ and NB1261 filters.
We assume [\ion{N}{2}] contribution is 30 per cent \citep{1999MNRAS.310..262T}. 
We adopt the amount of dust extinction computed by MIZUKI code (SED fitting) and listed in the HSC-SSP catalog, converted to {\Ha} extinction using the extinction law of \citet{2000ApJ...533..682C}.
Note that we assume the color excess of nebular gas emission is equal to that of stellar continuum.
We then calculate star formation rates from the SFR-$L_{\mathrm{H}\alpha}$ relation of \citet{1998ARA&A..36..189K}:
\begin{equation}
	\frac{\mathrm{SFR}}{M_{\sun} \, \mathrm{yr}^{-1}}
	= 7.9 \times 10^{-42} \frac{L_{\mathrm{H}\alpha}}{\mathrm{erg \, s^{-1}}}.
	\label{eq:SFR_Ha}
\end{equation}
IR luminosity is also convert to SFR by the relation of \cite{1998ARA&A..36..189K}:
\begin{equation}
	\frac{\mathrm{SFR}}{M_{\sun} \, \mathrm{yr}^{-1}} 
	= 4.5 \times 10^{-44} \frac{L_{\mathrm{IR}}}{\mathrm{erg \, s^{-1}}}.
	\label{eq:SFR_IR}
\end{equation}
The total IR luminosity ($L_{\mathrm{IR}}$) is estimated by \citet{2019MNRAS.484.4695T} from the Spitzer/MIPS data using the template of \citet{2008ApJ...682..985W}.
The two relations above have assumed Salpeter IMF \citep{1955ApJ...121..161S}; therefore in order to convert it to Chabrier IMF \citep{2003PASP..115..763C} we use the relation of \citet{2013MNRAS.430.2622D}.
In addition, we calculate SFR(FIR+UV) by the relation of \citet{2005ApJ...625...23B}:
\begin{equation}
	\frac{\mathrm{SFR}}{M_{\sun} \, \mathrm{yr}^{-1}} 
	= 2.5 \times 10^{-44} \frac{L_{\mathrm{IR}} + 2.2L_{\mathrm{UV}}}{\mathrm{erg \, s^{-1}}}
	\label{eq:SFR_IR_UV}
\end{equation}
where $L_{\mathrm{UV}} = 1.5 \nu l_{\nu,2800}$ and $l_{\nu,2800}$ is the rest-frame 2800 {\AA} luminosity density obtained from the SED fitting. 

\section{Results}
\subsection{Colors and Magnitudes}

\begin{figure*}
	\centering
	\includegraphics[width=\linewidth]{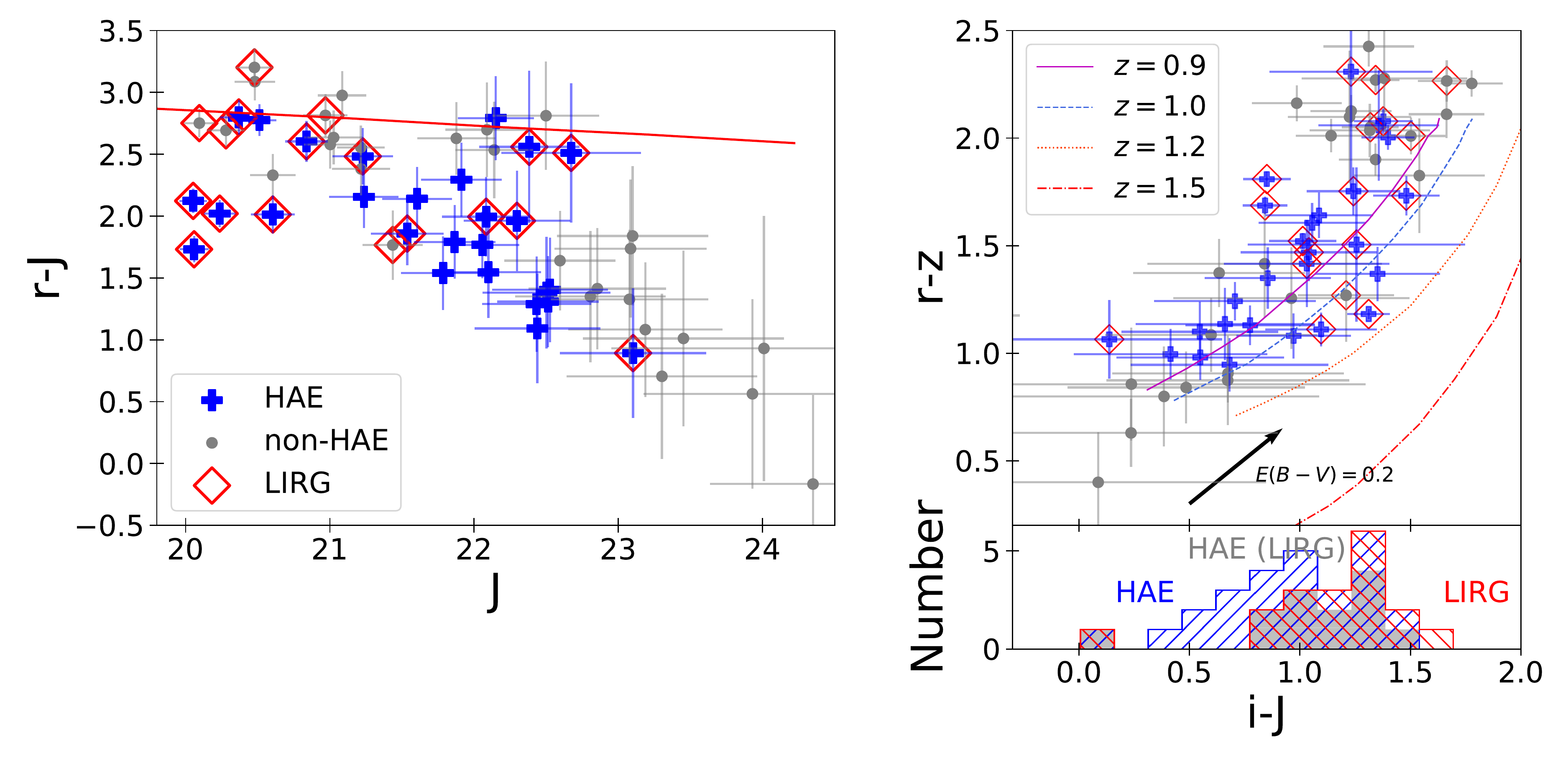}
	\caption{Left: $r-J$ versus $J$ color-magnitude diagram. Right: $r-z$ versus $i-J$ color-color diagram and $i-J$ color distribution. HAEs and non-HAEs are marked with blue crosses and gray points, respectively. Red open diamonds indicate LIRGs. Red solid line in the left panel represents the red sequence at $z=0.95$ predicted for passively evolving galaxies formed at $z=2$. This is based on the spectral evolution model of early-type galaxies in \citet{1998A&A...334...99K}, which is constructed to reproduce the color-magnitude relation of Coma cluster early-type galaxies. In the right-hand panel, we show a reddening vector calculated from the Calzetti extinction law \citep{2000ApJ...533..682C} with $E(B-V)=0.2$ and $R_V=3.1$. We plot predicted color tracks based on the \citet{1999MNRAS.302..152K} model, where bulge-to-total ratio is changed from 1 to 0 along the curves. In the right bottom panel, $i-J$ color distributions of HAEs (blue hatched) and LIRGs (red hatched) are shown. In addition, that of HAEs which are also LIRGs (denoted as HAE (LIRG) here) is represented as a gray filled histogram.\label{fig:cmd_ccd}}
\end{figure*}

The $r-J$ versus $J$ color-magnitude and $r-z$ versus $i-J$ color-color diagrams are shown in Figure~\ref{fig:cmd_ccd}. HAEs and cluster members which are not selected as HAEs (hereafter, we call them non-HAEs) are marked with blue crosses and gray points, respectively, and red open diamonds indicate LIRGs ($L_{\mathrm{IR}}>10^{11}L_{\sun}$). 
It is seen that the colors and $J$-band magnitudes of HAEs tend to be different depending on their IR luminosity. 
In fact, half of the HAEs are also LIRGs, but the others are not.
The HAEs which are also LIRGs seem to be more luminous and redder than the non-LIRG HAEs, and some of them are actually located on the red sequence ($r-J \gtrsim 2.5$). They are likely to be experiencing strong dust attenuation.
In order to check whether colors of the HAEs actually differ depending on their IR luminosities, we performed a Kolmogorov–Smirnov (K-S) test. However, the probability ($p$-value) that the $i-J$ color distribution of the LIRG HAEs and that of the non-LIRG HAEs are generated by the same parent distribution is 11.1 \%. 
We also did a K-S test for $r-J$ color distributions of HAEs and LIRGs in the same magnitude bin of $21.5<J<23$, but the $p$-value is 15.4 \%. Hence we do not see a statistically significant difference.
Interestingly, four LIRG HAEs lie at the bright end of the blue cloud. They are not so reddened although they are as luminous in IR as dusty starbursts.
Some LIRGs which are not HAEs are on the red sequence.  In these galaxies, dust extinction may be so strong that even {\Ha} emission line cannot escape from the star forming regions. 
Such highly attenuated starbursts are reported in other clusters and fields \citep[e.g.,][]{1999ApJ...525..609S, 2001ApJ...550..195P, 2009ApJ...693..140D, 2009ApJ...693..152O}.

\begin{figure}[htbp]
	\centering
	\includegraphics[width=\linewidth]{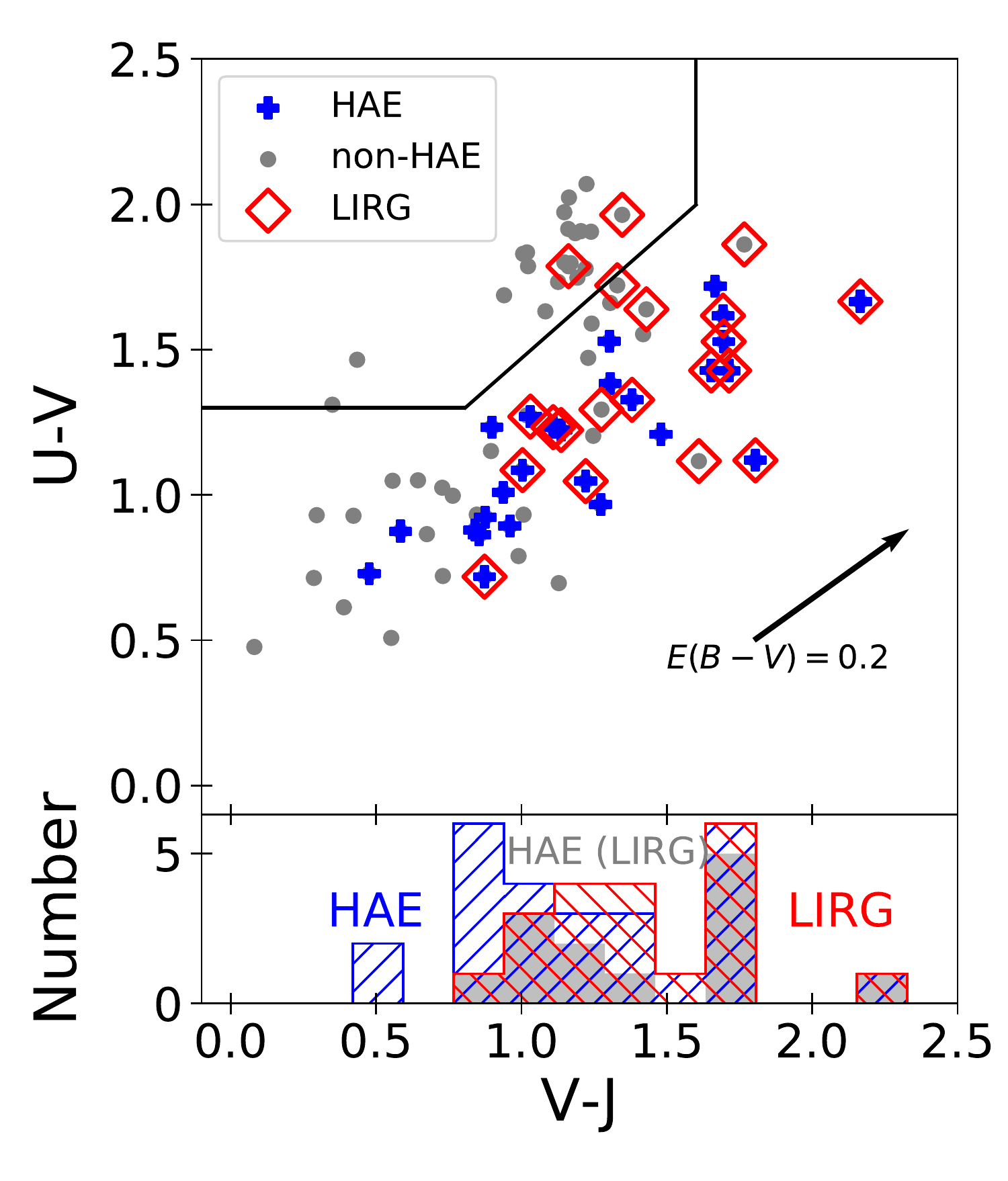}
	\caption{Rest-frame $UVJ$ diagram. We show the boundary from \cite{2009ApJ...691.1879W}, which separates quiescent galaxies from star-forming galaxies. Symbols are the same as those in Figure~\ref{fig:cmd_ccd}. The arrow indicates the reddening vector from the Calzetti extinction law. In the bottom panel, rest-frame $V-J$ color distributions of HAEs, LIRGs and HAEs (LIRG) are shown, respectively.\label{fig:uvj_diagram}}
\end{figure}

The same trends are seen on the rest-frame $U-V$ versus $V-J$ diagram ($UVJ$ diagram) in Figure~\ref{fig:uvj_diagram}. 
All the HAEs and most of the LIRGs (except for two) are classified as star-forming galaxies. 
The non-LIRG HAEs tend to be bluer.  In contrast, the LIRG HAEs are relatively reddened.
Here we performed K-S test again, and found that the probability that LIRG HAEs and non-LIRG HAEs have the same parent distribution in $V-J$ color is only 1.2 \%, indicating that LIRGs and HAEs trace the star forming galaxies in different phases.
LIRGs may trace starbursts with strong dust attenuation, while HAEs are likely to trace more normal star-forming galaxies with moderate dust extinctions.

\subsection{Star-forming Main Sequence}
\begin{figure*}[htbp]
	\gridline{
		\fig{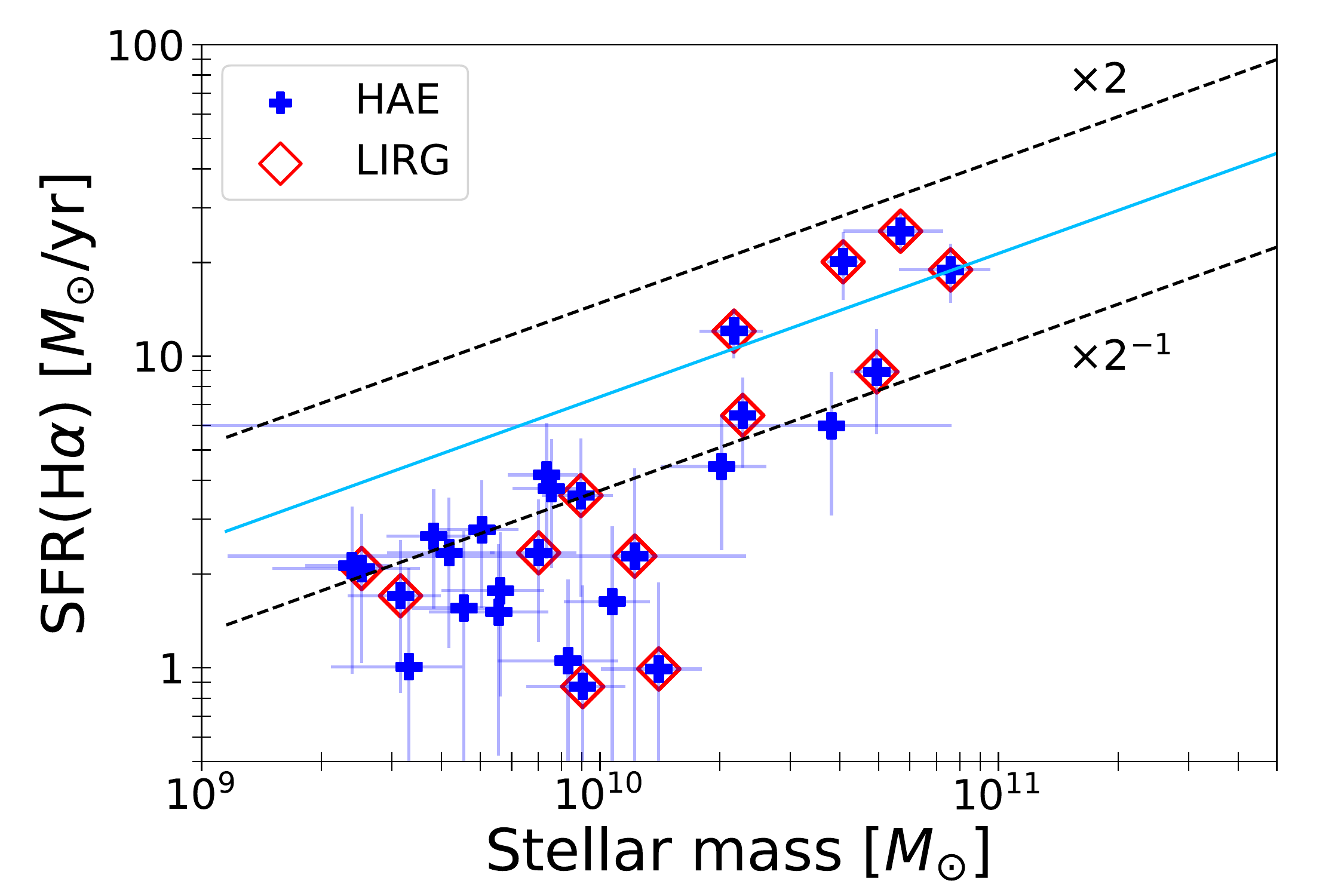}{0.5\linewidth}{}
		\fig{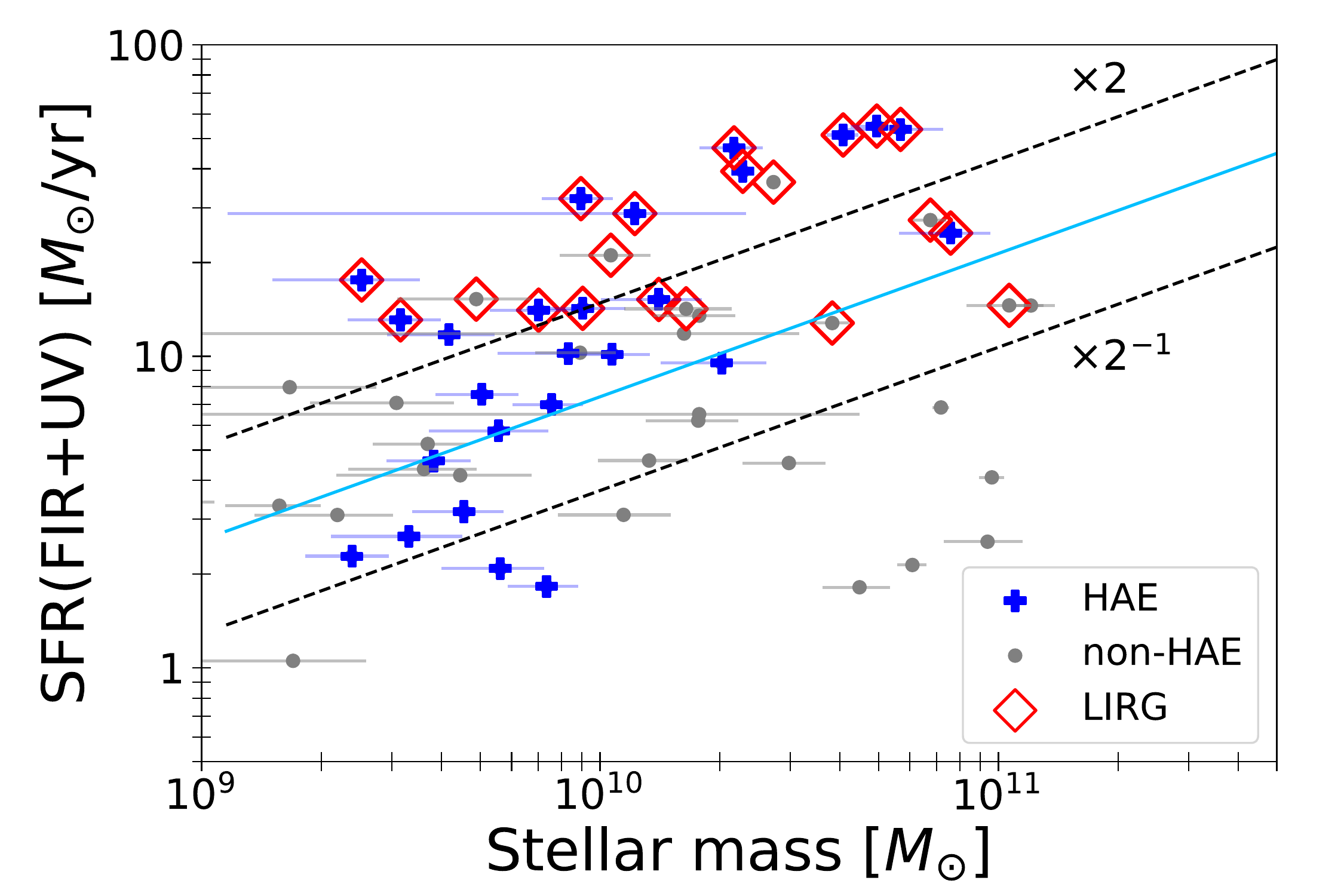}{0.5\linewidth}{}
	}
	\caption{Left: SFRs based on {\Ha} line luminosities and dust extinction correction as a function of stellar mass. Blue crosses and red diamonds represent HAEs and LIRGs, respectively.  We plot a main sequence of star-forming galaxies at $z=0.95$ from \citet{2018A&A...615A.146P} (light blue solid line). In addition, we plot a factor of 2 range above and below the main sequence by black dashed lines.
	Right: SFRs based on IR and UV luminosities. Gray filled circles represent non-HAEs. The other symbols and lines are the same as those in the left-hand panel.\label{fig:main_sequence_diagrams}}
\end{figure*}

We show {\Ha}-based and (FIR+UV)-based SFRs in the left and right panels of Figure \ref{fig:main_sequence_diagrams}, respectively.
We also plot the main sequence at $z \sim 0.95$ from \citet{2018A&A...615A.146P}, who estimate SFRs and stellar mass by a SED fitting for galaxies observed in multi-wavelengths (from UV to FIR).
They also assume Chabrier IMF \citep{2003PASP..115..763C}.
Low-mass HAEs whose stellar masses are smaller than $\sim 2 \times 10^{10} M_{\sun}$ have SFRs ({\Ha})  consistent with (or slightly lower than) the main sequence.
On the other hand, high-mass HAEs whose stellar masses are larger than $\sim 2 \times 10^{10} M_{\sun}$ have SFR({\Ha}) values that are comparable to the main sequence.

Significantly different trends are seen for SFR(FIR+UV) in the right-hand panel of Figure~\ref{fig:main_sequence_diagrams}. Most of the LIRGs, irrespective of their stellar mass, have higher SFR(FIR+UV) by a factor of two than those on the main sequence. In contrast, HAEs have SFR(FIR+UV) more or less consistent with the main sequence.
For LIRGs, however, {\Ha}-based SFRs tend to be even more underestimated than for non-LIRG HAEs even after the dust corrections based on the SED fitting. 

\subsection{Dust Extinction}
\begin{figure*}
	\gridline{
		\fig{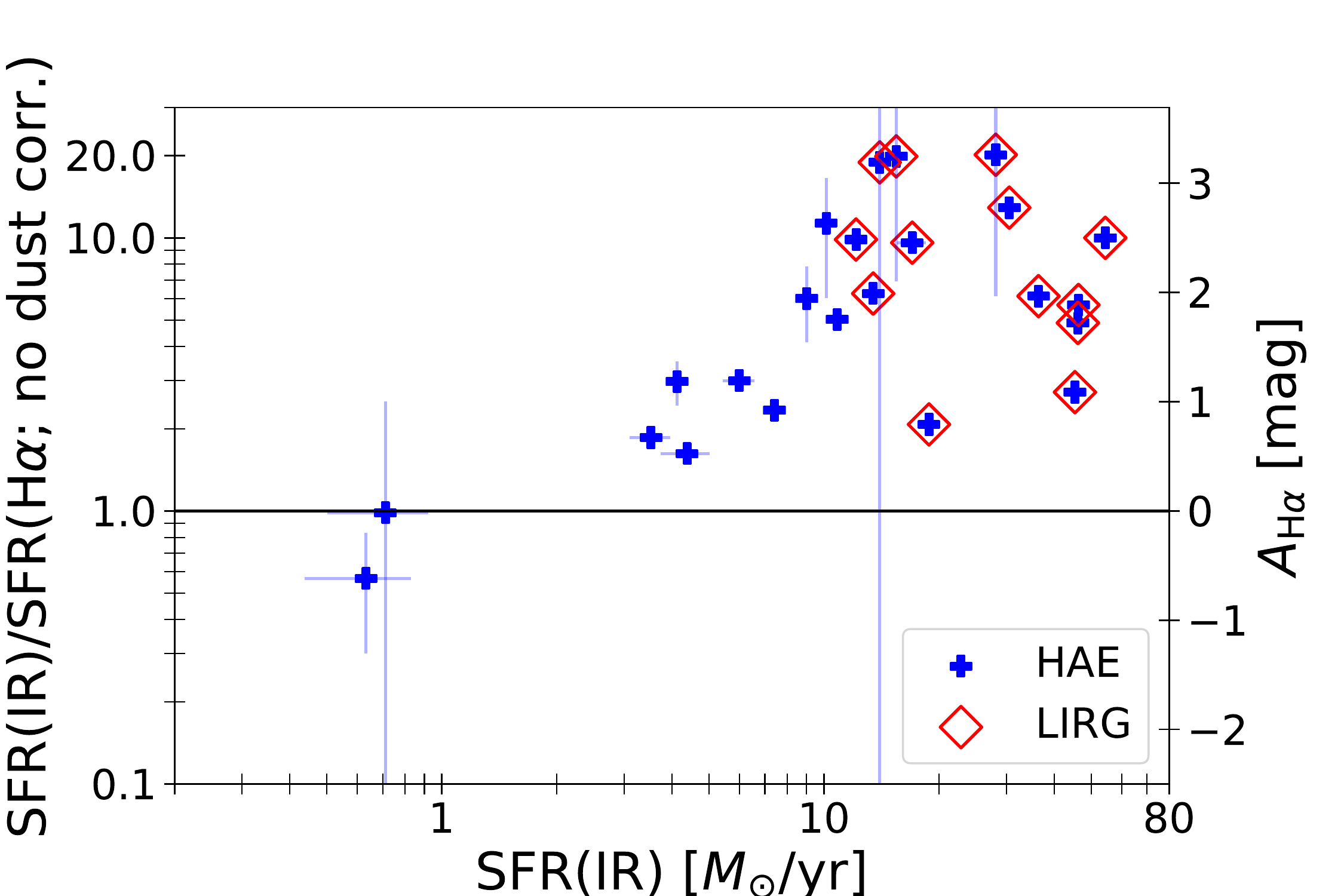}{0.5\linewidth}{}
		\fig{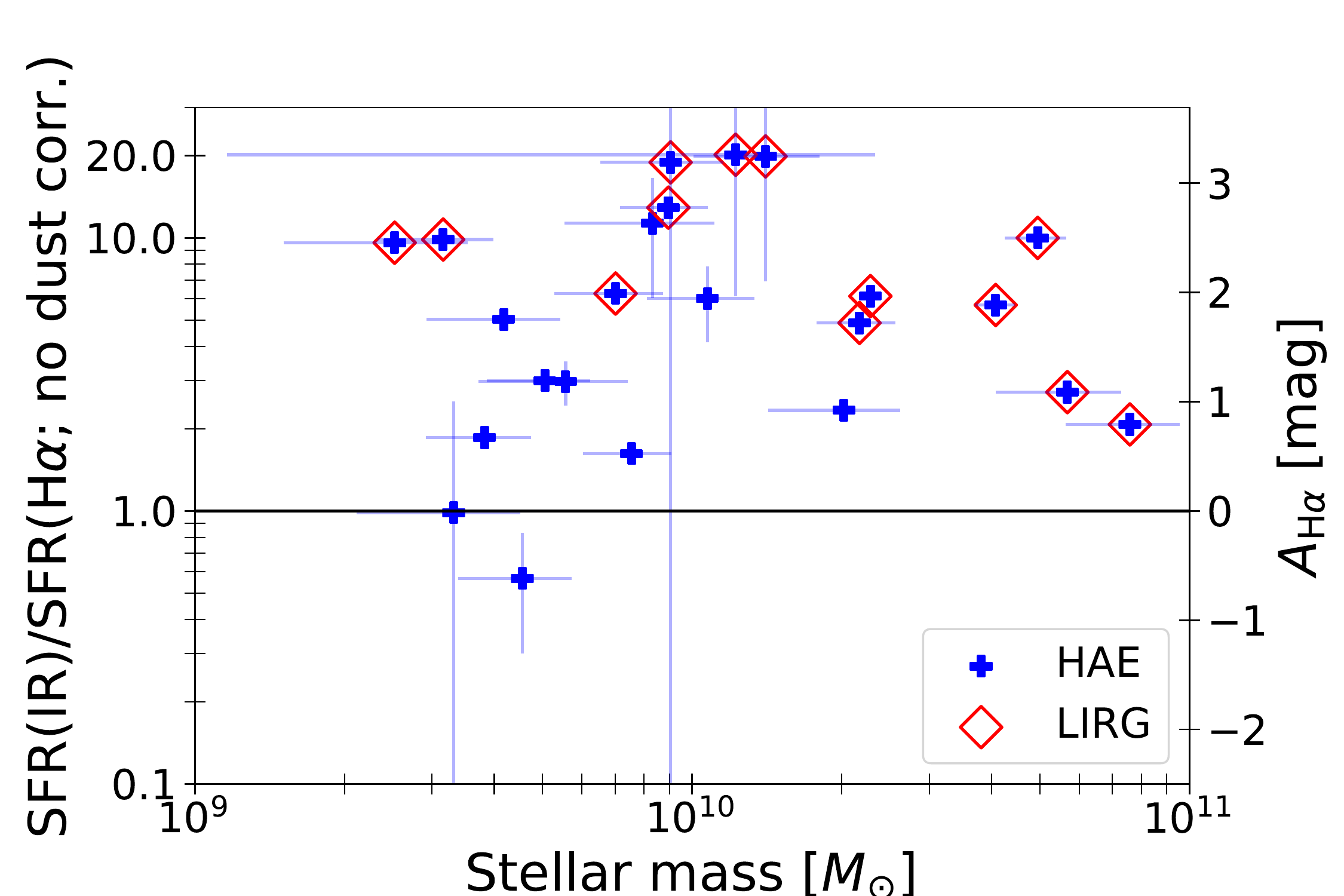}{0.5\linewidth}{}
	}
\caption{SFR(IR)/SFR({\Ha}; no dust corr.) as a function of SFR(IR) and stellar mass. Note that, in this diagram, the {\Ha}-based SFRs are not corrected for dust extinction unlike those in Fig.~\ref{fig:main_sequence_diagrams}. HAEs and LIRGs are marked with blue crosses and red diamonds respectively. }
	\label{fig:extinction}
\end{figure*}

Interstellar dust in star-forming galaxies mainly absorbs short wavelength radiation originated from young massive stars (which also emit {\Ha} line), and re-emit the energy as thermal radiation in IR wavelength.
Therefore SFR(IR)/SFR({\Ha; no dust corr.}) can be used as an index of dust extinction.
We show in Figure~\ref{fig:extinction} SFR(IR)/SFR({\Ha; no dust corr.}) as a function of SFR(IR) and stellar mass. 
Note that in this plot and in this section, {\Ha}-based SFRs are extinction uncorrected values.
In addition, we convert SFR(IR)/SFR({\Ha; no dust corr.}) to the extinction in magnitude at the {\Ha} wavelength, $A_{\mathrm{Ha}}$, by estimating the intrinsic {\Ha} flux without extinction from SFR(IR).
As we can see in the left-hand panel of Figure~\ref{fig:extinction}, SFR(IR)/SFR({\Ha; no dust corr.}) weakly increases with SFR(IR). 
This result that high-SFR galaxies tend to have high SFR(IR)/SFR({\Ha; no dust corr.}) values is consistent with the fact that they are dusty starbursts.
We confirm that this trend is not caused by a selection effect by the fact that that SFR(IR)/SFR({\Ha; no dust corr.}) does not correlate with SFR({\Ha}) instead.
The same trend is reported for star-forming galaxies in a cluster at $z \sim 0.8$ \citep{2010MNRAS.403.1611K}. 
\citet{2011ApJ...736...38K} show that in SC1604 with {\Oii} instead of {\Ha}.

On the other hand, according to right-hand panel of Figure~\ref{fig:extinction}, dust extinction depends little on stellar mass.
For field galaxies, it is shown that the amount of extinction tends to increase with stellar mass \citep{2010MNRAS.409..421G}.
The scatter in our sample is much larger than the original scatter of the relation in \citet{2010MNRAS.409..421G} ($\sim 0.3$ mag).
This extra scatter and the lack of mass dependence may indicate that some additional environmental effects may weaken the underlying intrinsic mass dependence of dust extinction as a function of stellar mass.
Possible candidates for such environmental effects are galaxy interactions and mergers, which enhance star formation in clusters.

\subsection{Structure of CL1604-D}\label{sec:structure}
\subsubsection{Spatial Distribution}\label{sec:spatial_distribution}
\begin{figure}
	\centering
	\includegraphics[width=\linewidth]{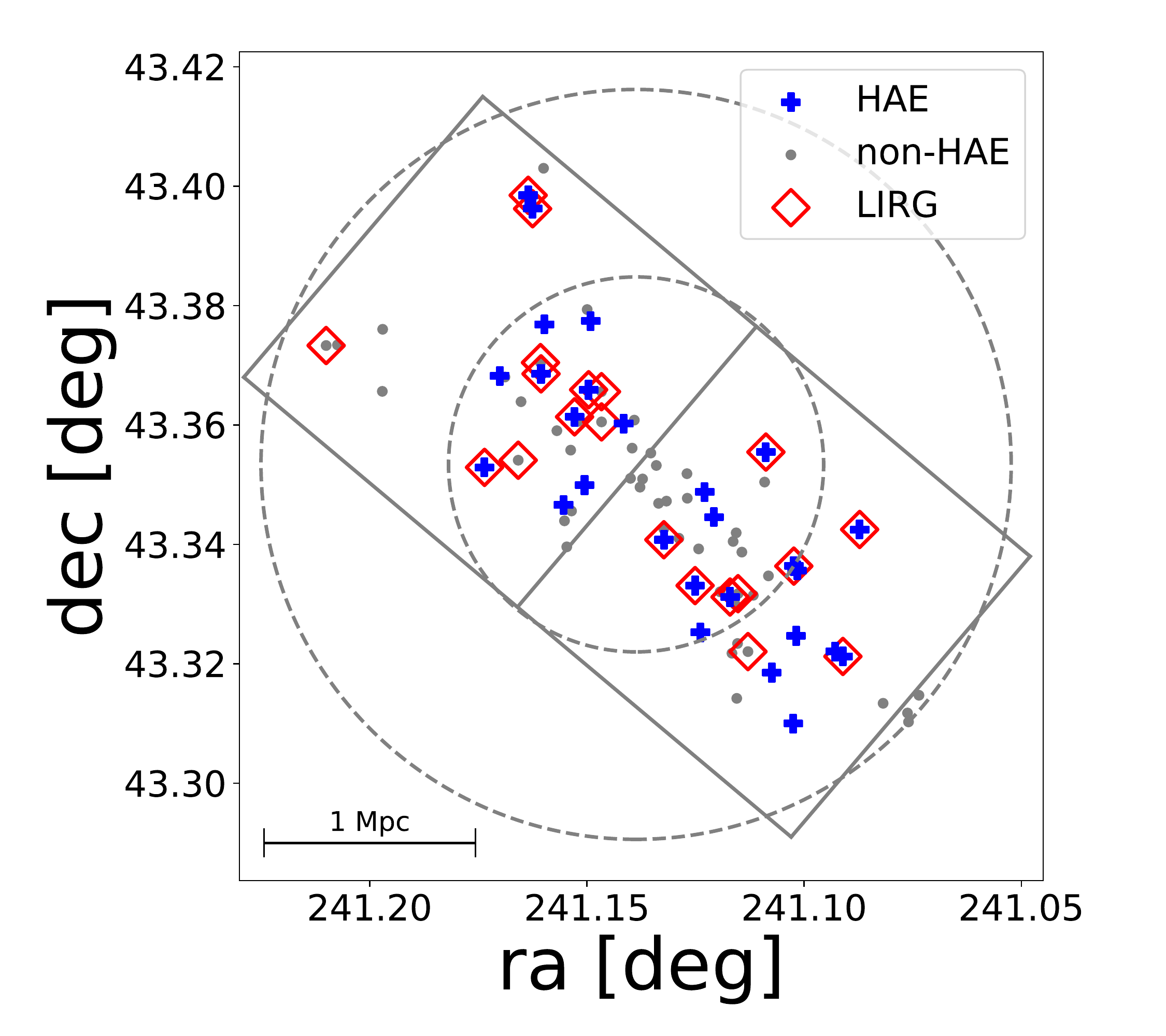}
	\caption{Spatial distribution of galaxies in and around the cluster CL1604-D. The symbols are same as Figure~\ref{fig:cmd_ccd}. The two squares indicate the FoV's of SWIMS detectors with two arrays. The two circles denote the virial radii of the cluster and the twice this size, respectively.\label{fig:spatial_distribution}}
\end{figure}

A spatial distribution of galaxies in and around the cluster CL1604-D is shown in Figure~\ref{fig:spatial_distribution}. Symbols in the Figure are same as those in Figure~\ref{fig:cmd_ccd}. 
Star-forming galaxies (both the LIRGs and the HAEs) are distributed along the filament, but they both avoid the cluster central region ($r \lesssim 0.3 \rvir$). Star-forming activity seems to be fully quenched there. On the other hand, many star-forming galaxies reside in outer northeast and southwest regions which are $0.3\rvir$ or more away from the center of the cluster. The northeast region is crowded with LIRGs, which make up a small group-like structure. Hereafter we call it a NE group (but qualitatively defined in \S\ref{sec:env_depend}). In the southwest region both HAEs and LIRGs form  a small clump. 
Hereafter we call it a SW group.
Note that we use the term of ``group'' in this study as a spatially clumped system consisting of $\sim10$ galaxies.
One possibility is that the two groups are in-falling systems into the cluster core. In the next section, we discuss kinematic properties of those systems and galaxies therein.

\subsubsection{Phase-Space Diagram}\label{sec:phase_space}
\begin{figure*}
	\plotone{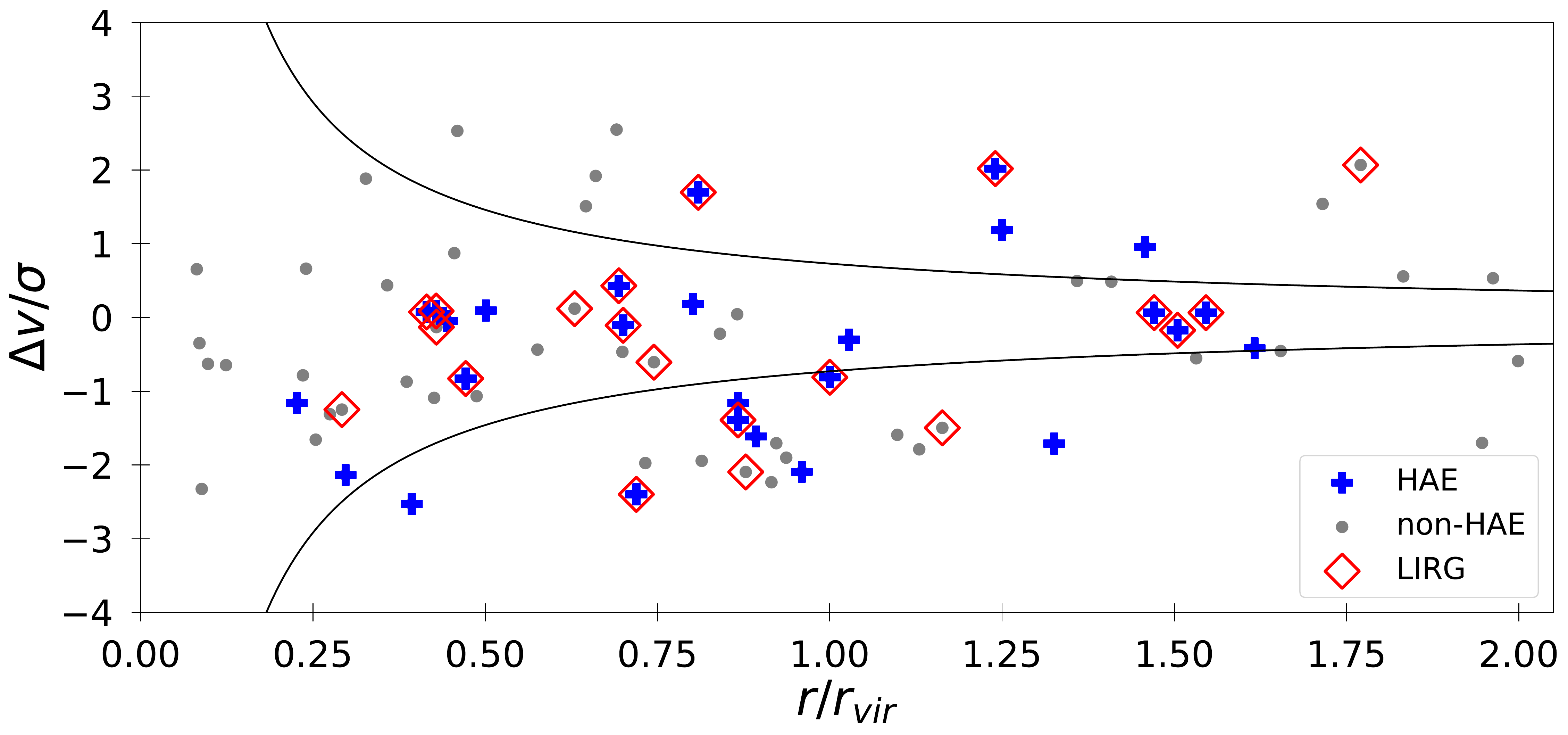}
	\caption{A phase-space diagram of the cluster. Radial velocity ($\Delta v$) divided by the line-of-sight velocity dispersion ($\sigma$) is plotted against the projected clustocentric distance ($r$) divided by the virial radius ($\rvir$).
Two curves roughly separate the phase space into a central/intermediate region and a recently according/in-falling region \citep{2013ApJ...768..118N, 2016ApJ...816...48N}. Symbols are same as those in Figure~\ref{fig:cmd_ccd}.\label{fig:phase_space_diagram}}
\end{figure*}

Here we discuss a kinematic structure of the cluster from a phase-space diagram, namely, $\Delta v/\sigma$ versus $r/\rvir$ plot where $\Delta v$ is a radial velocity offset from the median redshift of the cluster, $r$ a distance from the cluster center, $\sigma$ a radial velocity dispersion, and $\rvir$ a projected virial radius.
We use the phase-space analysis described in \citet{2013ApJ...768..118N, 2016ApJ...816...48N}, which divide the phase space into four regimes, central, intermediate, recently accreted, and in-falling regions.
In a relaxed system, velocity dispersion of galaxies in the cluster core is larger than that in the outskirts of the cluster, and the virialized population should lie roughly in the central or intermediate regions. In contrast, in-falling groups have velocity offsets and tend to appear separate from the virialized populations.

We show in Figure~\ref{fig:phase_space_diagram} the phase-space diagram of the galaxies in and around the CL1604-D cluster. Two curves (corresponding to $(r/r_{200}) \times (\Delta v/\sigma) = \pm 0.64$) in the figure show boundaries in the phase space which separate the intermediate and the recently accreted regions. We convert $r_{200}$ to $\rvir$ by the relation: $\rvir = r_{200}/1.14$ \citep{2006A&A...456...23B, 2009ApJ...693..112P}.
We can see a kinematically distinct and in-falling group below the lower curve in the figure.
It corresponds to a part of the SW group in Figure~\ref{fig:spatial_distribution} (although connections between clumps in the phase space and those in the projected spatial map is discussed in \S\ref{sec:env_depend}). 
Approximately half of the galaxies in this group are HAEs and/or LIRGs. 

On the other hand, member galaxies in the NE group are crowded also in the phase-space diagram. They reside at $r/\rvir \sim 0.5-1$ and $ | \Delta v/ \sigma| \lesssim 1 $. They lie between the two curves in Figure~\ref{fig:phase_space_diagram} and seem to be virialized members from the phase-space diagnosis alone. However, the NE group is very likely an in-falling group because they share similar line-of-sight velocities and spatial positions. 
They are moving toward the cluster core probably almost perpendicular to the line of sight, so that their line-of-sight velocity may not be large.

\subsection{Environmental Dependence}\label{sec:env_depend}
\begin{figure*}
	\centering
	\includegraphics[width=\linewidth]{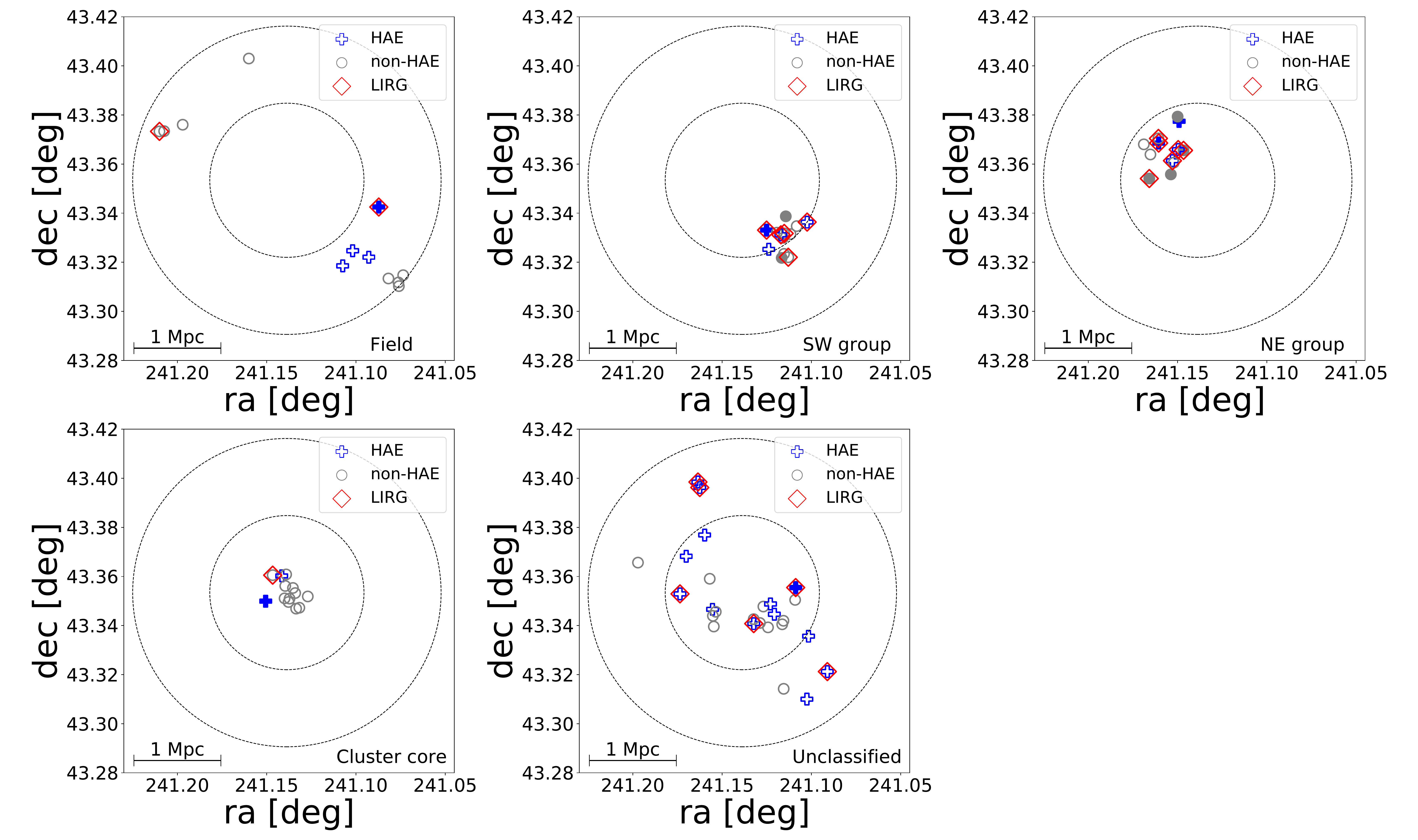}
	\caption{Spatial distributions of galaxies in each of the environments. HAEs and non-HAEs are marked with blue crosses and gray circles, respectively. Red open diamonds indicate LIRGs. Filled symbols represent mergers or interacting galaxies\label{fig:spatial_distribution_sub}}
\end{figure*}

\begin{figure*}
	\centering
	\includegraphics[width=\linewidth]{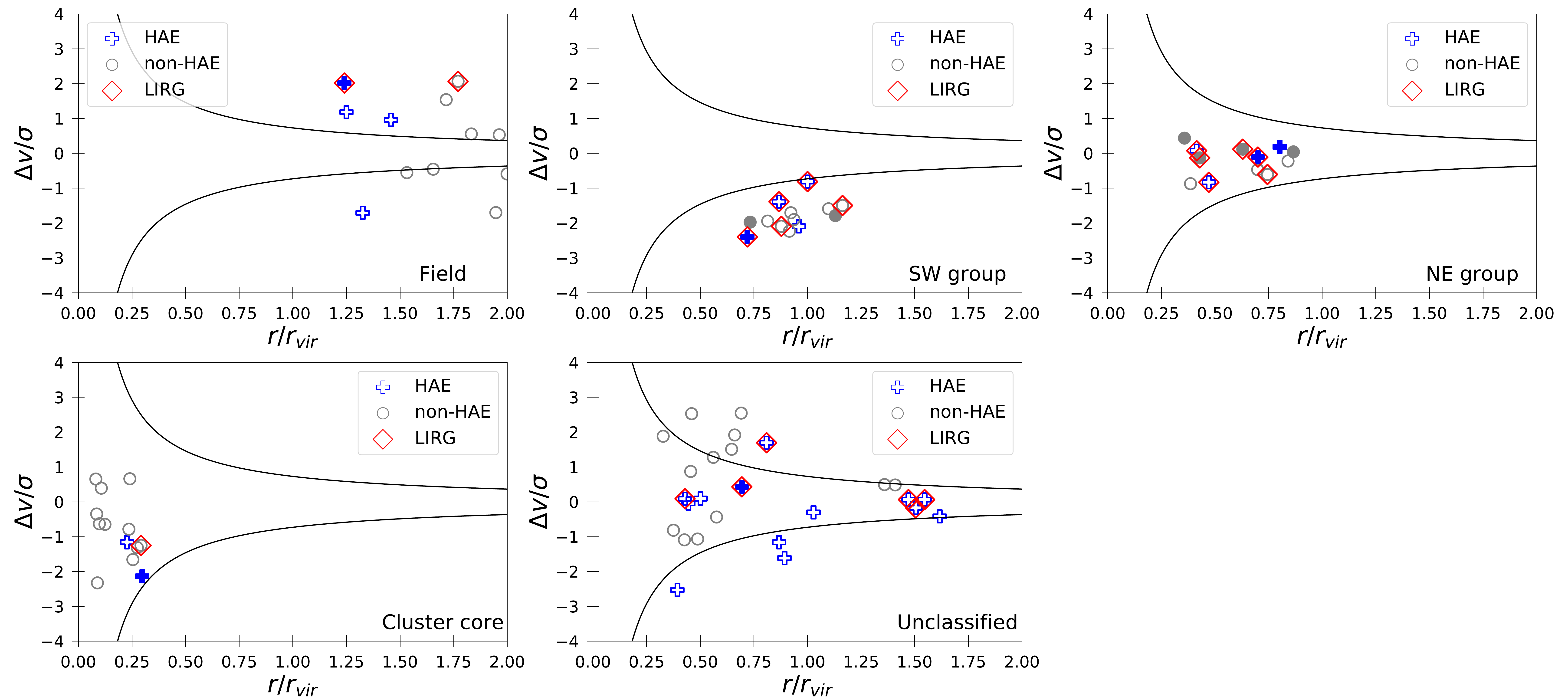}
	\caption{Phase-space diagrams of the galaxies in each of the environments. Symbols are the same as those in Figure~\ref{fig:spatial_distribution_sub}.\label{fig:phase_space_diagram_sub}}
\end{figure*}

We qualitatively define four specific environments on the following conditions according to the substructures/groups of galaxies:
(1) cluster core; galaxies that satisfy $r/\rvir < 0.3$ and $|(r/r_{200})\times (\Delta v/\sigma)| < 0.63$,
(2) NE group; galaxies that reside in the northeast side of the cluster (which corresponds to  the array-1), and satisfy $0.3 < r/\rvir < 1$ and $|\Delta v / \sigma | < 1$,
(3) SW group; galaxies that reside in the southwest side of the cluster (which corresponds to the array-2), and satisfy $0.7 < r/\rvir < 1.25$ and $ (r/r_{200}) \times (\Delta v/\sigma) < - 0.64$, and
(4) field: galaxies that satisfy $r/\rvir > 1$ and $|(r/r_{200})\times (\Delta v/\sigma)| > 0.63$ but do not belong to the NE or SW group.
Spatial distribution and phase-space diagram for each substructure are shown in Figure~\ref{fig:spatial_distribution_sub} and Figure~\ref{fig:phase_space_diagram_sub}, respectively.
We also show the same plots for galaxies which are not classified into any of the four environments for comparison (called as "unclassified"). We do not use them in the analyses of environmental dependence because it is a miscellaneous class and unclear which environment they belong to based on their 2D sky distribution or in the phase space diagram.
The combination of 2D spatial and kinematic information enables us to define the well-separated four environments. The velocity dispersion of the cluster core is $\sim  670\; \mathrm{km\;s^{-1}}$ which is comparable to the cluster's global value. 
However, the velocity dispersions of both SW group and NE group are $\sim 280 \; \mathrm{km\;s^{-1}}$. This relatively smaller velocity dispersion also suggests that these are less massive systems, which can be called as "groups".

\begin{deluxetable*}{ccccccccc}
	\tablecaption{Star-forming and morphological properties of the cluster members.\label{tab:star-foming_morph}}
	\tablehead{
		\colhead{Environments}  & \colhead{Number} & \colhead{HAE} & \colhead{LIRG} 
								& \colhead{Starburst} & \colhead{Merger/Interaction} 
								& \colhead{} & \colhead{$\mssfr$} & \colhead{}\\
		\colhead{}  & \colhead{} & \colhead{} & \colhead{} 
								& \colhead{(\%)} & \colhead{(\%)} 
								& \colhead{} & \colhead{($10^{-10}\mathrm{yr}^{-1}$)}  & \colhead{}\\
								\cline{7-9}
		\colhead{}  & \colhead{} & \colhead{} & \colhead{} 
								& \colhead{} & \colhead{} 
								& \colhead{Starburst} & \colhead{Non-starburst} & \colhead{total}
	}
	\startdata
	Field & 12 & 4 & 2  & $25^{+16}_{-8.3}$ & $8.3^{+15}_{-2.8}$ & 19 & 1.6 & 5.5 \\ 
	SW group & 13 & 4 & 5 & $54^{+12}_{-13}$ & $23^{+15}_{-7.6}$ & 15 & 8.7 & 15 \\ 
	NE group & 12 & 4 & 6 & $17^{+16}_{-5.9}$ & $50^{+13}_{-13}$ & 15 & 3.4 & 5.5 \\ 
	Groups (SW+NE) & 25 & 8 & 11 & $36^{+10}_{-8.2}$ & $36^{+10}_{-8.2}$ & 15 & 3.7 & 8.0 \\ 
	Cluster core & 13 & 2 & 1 & $0^{+12}_{-0}$ & $7.7^{+14}_{-2.5}$ & --- & 1.1 & 1.1 \\
	\hline
	Unclassified & 26 & 13 & 6 & $38^{+9.9}_{-5.5}$ & $3.8^{+7.8}_{-1.2}$ & 16 & 2.2 & 4.3 
	\enddata
\end{deluxetable*}

\subsubsection{Star Formation Activity}\label{sec:sf_env}

\begin{figure*}
	\centering
	\includegraphics[width=\linewidth]{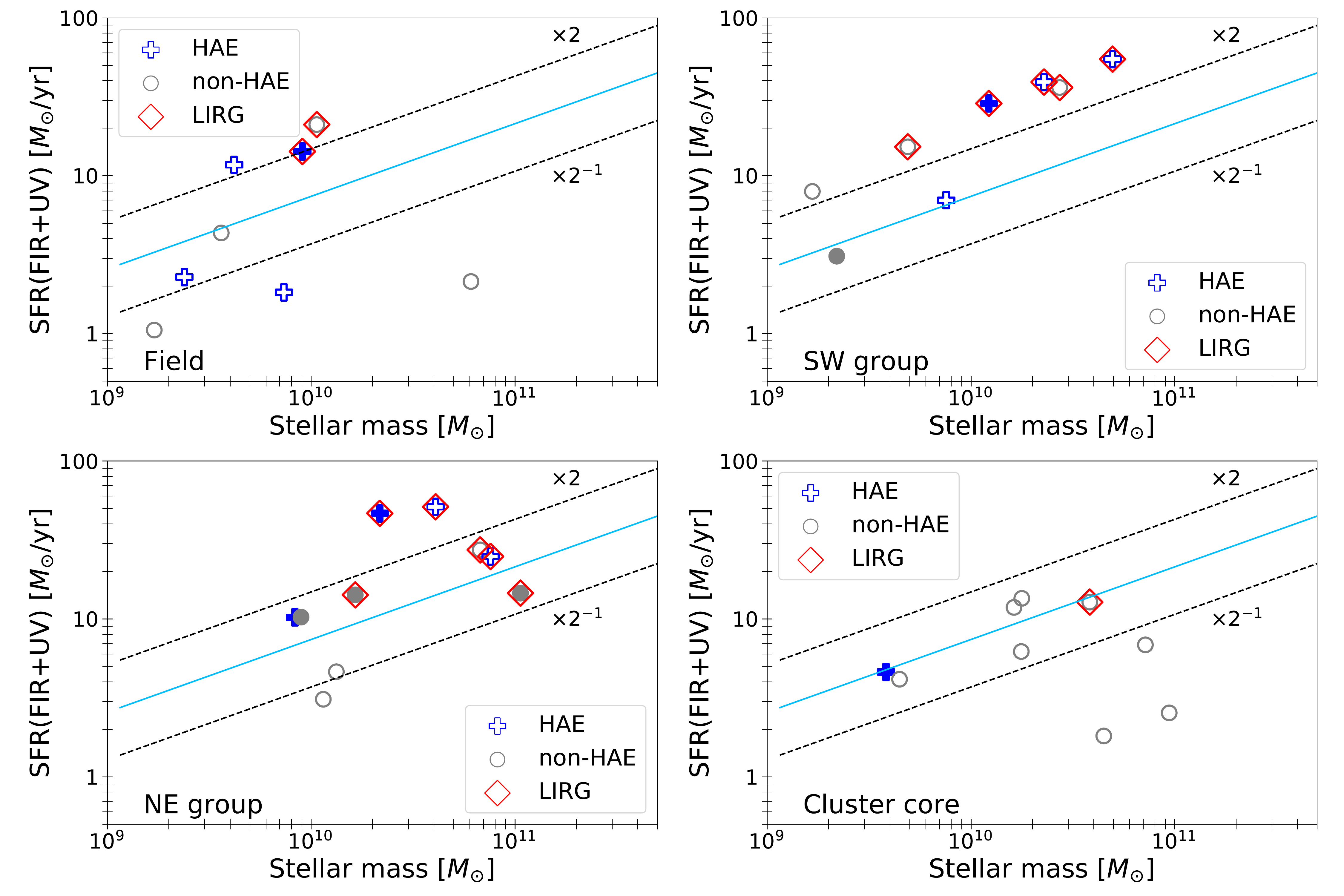}
	\caption{Main-sequence diagram of the galaxies in each of the environments. Symbols are the same as those in Figure~\ref{fig:spatial_distribution_sub}.\label{fig:main_sequence_diagrams_sub}}
\end{figure*}

The main-sequence diagram for each region is shown in Figure~\ref{fig:main_sequence_diagrams_sub}.
In order to quantify the burstiness of star formation, we define mean specific star formation rate ($\mssfr$) by the following equation:
\begin{equation}
	\mssfr 
	=\frac{\sum \mathrm{SFR(FIR+UV)}}{\sum M_{\star}}
\end{equation}
where $\sum \mathrm{SFR(FIR+UV)}$ and $\sum M_{\star}$ are total SFR and total stellar mass in each environment, respectively. In addition, we define starbursts as the galaxies whose SFR(FIR+UV) are higher than that of the main sequence by a factor of two. 
We show the starburst fraction and $\mssfr$ of all the galaxies as well as those of starbursts and non-starbursts in each system in Table~\ref{tab:star-foming_morph}.
Due to the "color-term" in spectroscopy (color dependence in the successful redshift determination), the stellar mass limit of red galaxies is higher than that of blue galaxies, hence starburst fraction and $\mssfr$ may not be exact values in absolute sense. However, we can still do a fair comparison among different environments in a relative sense, because the effects of the color-term work in the same way in all the environments.

Mean SSFRs vary from region to region. In the cluster core, which is the densest region in the structure, $\mssfr$ is lower than in any other regions. Furthermore, there is no starbursting galaxy.
In the NE group and the field, moderate star-forming activities are seen. In the SW group, half of the members are starbursts, and $\mssfr$ is roughly three times higher than that in the NE group or the field.
The difference of star formation burstiness between the groups is determined not by the star formation activity of individual galaxies, but by the fraction of starbursts in each group.

\begin{figure}
    \centering
    \includegraphics[width=\linewidth]{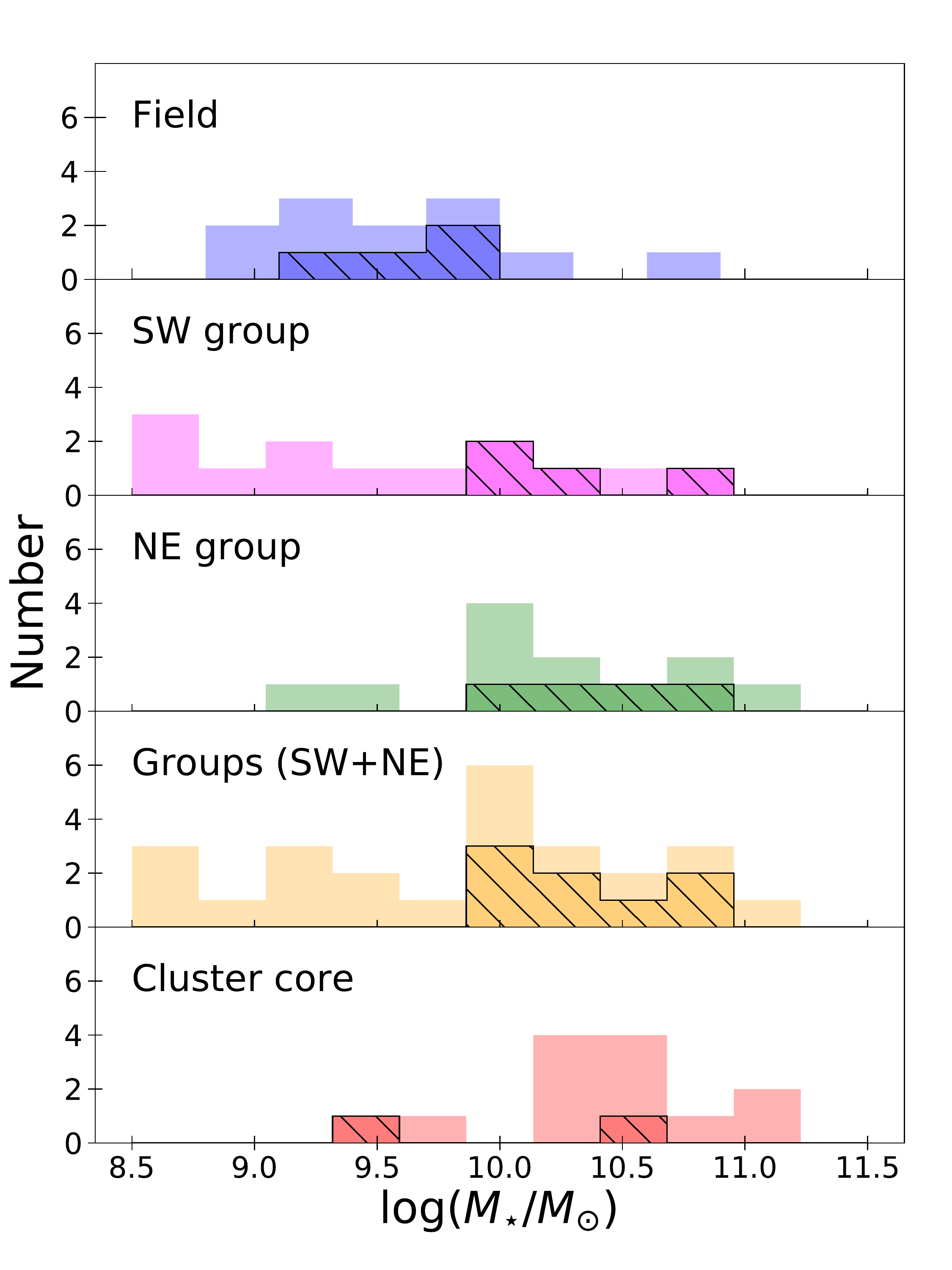}
    \caption{Stellar mass distribution in each substructure. Filled and hatched histograms correspond to all the galaxies and HAEs in each environment, respectively.\label{fig:mass_hist}}
\end{figure}

In Figure~\ref{fig:mass_hist}, we show stellar mass distributions in each environment, where histograms for all the galaxies (filled) and for HAEs only (hatched) in respective regions are shown. 
Stellar masses of galaxies in the field fall in the intermediate mass range ($9 \lesssim \log(M_{\star}/M_{\sun}) \lesssim 10.5$). The SW group galaxies are distributed in the widest mass range.  
The NE group and the cluster core have similar mass distributions. 
We note that the most massive galaxies ($\log(M_{\star}/M_{\sun})>11$) are seen only in these two regions. 
We see that the mass distribution is shifted from low to high mass direction in the order of the panels (from the top to bottom) in Figure~\ref{fig:mass_hist} where the galaxy
number density increases roughly in this order. 
K-S tests show significantly differences in mass distribution for pairs of field-NE ($p=3.1$\%), field-core ($p=0.05$\%), SW-NE ($p=3.9$\%) and SW-core ($p=1.2$\%).
However, if we limit our samples to HAEs, we cannot confirm mass distribution differences.

\subsubsection{Merger Fraction}\label{sec:merger}
In this study, we select mergers/interacting galaxies from the previous study \citep{2011ApJ...736...38K} as those identified as merger, interaction or tidal features.
In Figures~\ref{fig:spatial_distribution_sub}, \ref{fig:phase_space_diagram_sub} and \ref{fig:main_sequence_diagrams_sub}, the merger/interacting galaxies (hereafter referred to as merger galaxies for brevity) are marked with filled symbols, and their fractions in each region are listed in Table~\ref{tab:star-foming_morph}.
Similar to the star formation activities, the merger fractions largely differ from region to region.
In the field, only one merger is found. The density would be too low for the galaxies to interact with other galaxies gravitationally.
Mergers do not occur in the cluster core either, even though it is the densest region of the structure. This is probably because relative velocities among the galaxies in cluster core are very high due to the deep gravitational potential well, and therefore although the galaxies frequently encounter with others they just pass by and do not merge together.
As shown in \citet{2011ApJ...736...38K}, in the SW group and the NE group, a large number of LIRGs are located and the merger fraction is also high. In fact, the merger fraction in the SW and NE group is $\sim20$\% and 50\%, respectively, compared to $\sim$8\% in the field. 
We note however that, as shown in \S\ref{sec:sf_env}, we see $\mssfr$ is about three times higher in the SW group than in the NE group. 
There might be a time lag between mergers/interactions and induced star formations.
$N$-body/hydrodynamical simulations \citep[e.g.,][]{2008MNRAS.391.1137L} suggest starbursts induced by mergers are delayed from morphological disturbances.

\section{Discussion}
As shown in the previous sections, star formation and merger activities strongly depend on environments in the structure. In this section, we discuss relationship between these activities and history of galaxy accretion.
We defined four environments in and around the CL1604-D (field, SW group, NE group and cluster core) from their spatial and phase-space distributions, and the definitions of these regions seem to correspond to different accretion phases. The field galaxies are distinct from those in the cluster, which are defined by their  distance from the cluster center and their relatively large velocity offsets. They show moderate star formation activity, and their stellar masses tend to be lower than in other regions. Three of twelve galaxies are non-LIRG HAEs. On the contrary, the cluster core consists of galaxies which are assembled at the earliest times.
The high velocity dispersion of galaxies in this region implies that the system is already dynamically relaxed.
In addition, its low $\mssfr$ and high stellar mass are consistent with the fact that they are old populations which have already quenched their star formation activity.

The SW group and the NE group are defined as groups of galaxies which are spatially clustered and share similar velocity offsets.
Galaxies in the two groups are likely to be in-falling to the cluster core, still keeping their structures as groups.  Most of LIRGs reside in either of the two groups, and high star formation activities are seen there.
Another prominent feature of the two groups is that there are many mergers/interacting galaxies, which are not frequently seen in the field or in the cluster core.
These results suggest that the in-falling process and galaxy interactions trigger the starburst activity. This is already mentioned in previous studies such as \citet{2011ApJ...736...38K}. However, comparing $\mssfr$ and merger/interaction fraction in the groups, the merger activity does not seem to be completely linked to the star formation activity; $\mssfr$ is higher in the SW group than in the NE group although the merger/interaction fraction shows the opposite trend. 
One scenario is that the NE group accreted onto the cluster earlier than the SW group. As a result, its star formation activity was subsided and will be eventually quenched earlier than that in the SW group. 
In contrast, if we assume morphological transformations by galaxy mergers occur before the induced starbursts, galaxies in the SW group look relatively younger and may correspond to the pre-merger dominated phase. Their starbursts may be triggered even more in the near future.

In summary, properties of star forming galaxies seem to be correlated with the environments that we split based on the spatial and kinematic information, corresponding to the different phases of accretion history.
In fact, mergers and starbursts seem to be triggered during the course of accretion to the cluster, and then eventually fully quenched by the time they are incorporated into the cluster core populations.
High density in-falling groups are preferred environments for galaxy mergers to occur, and star formation activities therein are first enhanced and then quenched.
On the other hand, galaxies in the cluster core have experienced these processes a long time ago, and have been almost fully quenched by $z\sim1$. Velocity dispersion is too high for galaxies to actually merge together there, and instead ram-pressure stripping of associated gas in the in-falling galaxies by the inter-cluster medium (ICM) would terminate any remaining activities there.
This scenario agree with $N$-body simulations \citep[e.g,][]{2003ApJ...582..141G},  which show high-speed encounters in central regions of clusters rarely lead galaxies to merge.
We note that there is a small offset between a center of X-ray emission and a luminosity/mass-weighted center in CL1604-D \citep{2018MNRAS.478.1403R}.

The combination of the wealth of existing data available for the SC1604 supercluster and the new SWIMS/HSC wide-field imaging data has enabled us to investigate the star-forming activities as a function of substructures in and around the cluster in a fair amount of detail.
The existing MIPS 24 {\micron} imaging data have already provided information about star formation activities across the supercluster, but they trace only the dustiest phase of star formation, and our new {\Ha} emission line observations can make a very nice complimentary contribution.
As {\Ha} emission line is an excellent tracer of OB stars and less attenuated by dust compared to UV, we are able to map and detect weakly star-forming galaxies, or the normal phase of star formation by the narrow-band {\Ha} imaging technique.
In very dusty conditions such as a starbursting phase, however, even {\Ha} can no longer penetrate through the dusty clouds, and this phase can be complemented by the IR observations.
Both the IR and {\Ha} methods have pros and cons.
In fact, as seen in Figure~\ref{fig:cmd_ccd}, $\sim50$ \% of the HAEs are not LIRGs, and $\sim30$ \% of the LIRGs are not detected in {\Ha}.
Therefore the IR-{\Ha} combination is necessary in order to cover galaxies with various star-forming modes (from weak star formation to dust hidden starburst) and to obtain more comprehensive star formation histories in galaxies with little bias.

\section{Conclusions}
We have conducted a narrow-band {\Ha} imaging survey for the CL1604-D cluster at $z=0.923$ using a new infrared instrument SWIMS on the Subaru Telescope. Combined with existing data such as Subaru HSC optical imaging, HST imaging, intensive spectroscopic data and Spitzer MIPS imaging, we have obtained the following results.
\begin{enumerate}
\item We have identified 27 HAEs in the cluster, and they are classified as star-forming galaxies on the $UVJ$ diagram. Half of the HAEs overlap with LIRGs and are more luminous and redder than non-LIRG HAEs. 
		Bluer HAEs trace weak star-forming galaxies whose dust emissions are too faint to be detected in the IR.\@ {\Ha} emissions from some LIRGs are not detected, and they should be strongly dust attenuated star-forming galaxies.

	\item SFR(IR)/SFR({\Ha; no dust corr.}) as a measure of extinction weakly increases with SFR(IR), which is reported for a cluster at $z \sim 0.8$ \citep{2010MNRAS.403.1611K}. On the other hand, a correlation between dust extinction and stellar mass, which is typical for fields galaxies \citep{2010MNRAS.409..421G}, is not seen. Additional environmental effects such as galaxy interactions and mergers may have weakened the underlying intrinsic mass dependence on dust extinction.		

	\item We defined four environments in and around the cluster (field, SW group, NE group and cluster core) from their spatial and phase-space distributions. 
		In two in-falling groups, especially in the SW group, mean specific star formation rates are higher than in the field or the core. In addition, merger/interacting galaxies prefer the group environment. These results suggest that galaxy interactions trigger starburst activity.
	On the contrary, in the cluster core, starburst galaxies do not reside. Star formation seems to be quenched there.
\end{enumerate}

\acknowledgements
TK acknowledges support by JSPS KAKENHI Grant Number JP18H03717.
This work is supported by Ministry of Education, Culture, Sports, Science and Technology of Japan, Grant-in-Aid for Scientic Research
(18H03717, 15H02062, 23540261, 24103003, 24244015, 2611460, and 266780) from the JSPS of Japan.
Development of SWIMS is funded by a supplementary budget for economic stimulus packages formulated by the Japanese government.
This paper makes use of software developed for the Large Synoptic Survey Telescope. We thank the LSST Project for making their code available as free software at \url{http://dm.lsstcorp.org}. 
The Pan-STARRS1 Surveys (PS1) have been made possible through contributions of the Institute for Astronomy, the University of Hawaii, the Pan-STARRS Project Office, the Max-Planck Society and its participating institutes, the Max Planck Institute for Astronomy, Heidelberg and the Max Planck Institute for Extraterrestrial Physics, Garching, The Johns Hopkins University, Durham University, the University of Edinburgh, Queen's University Belfast, the Harvard-Smithsonian Center for Astrophysics, the Las Cumbres Observatory Global Telescope Network Incorporated, the National Central University of Taiwan, the Space Telescope Science Institute, the National Aeronautics and Space Administration under Grant No. NNX08AR22G issued through the Planetary Science Division of the NASA Science Mission Directorate, the National Science Foundation under Grant No. AST-1238877, the University of Maryland, and Eotvos Lorand University (ELTE).

\facilities{Subaru (HSC, SWIMS), Spitzer (MIPS), HST (ACS), Keck:I (LRIS), Keck:II (DEIMOS)}
\software{
	MCSRED2 (\url{http://www.naoj.org/staff/ichi/MCSRED/mcsred.html}),
	IRAF (\url{http://iraf.noao.edu}),
	SExtractor \citep{1996A&AS..117..393B},
	LSST Codes (\url{http://dm.lsstcorp.org})
}
\bibliographystyle{aasjournal}
\bibliography{reference}

\end{document}